\documentclass[fleqn,usenatbib]{mnras}

\usepackage{newtxtext,newtxmath}

\usepackage[T1]{fontenc}

\DeclareRobustCommand{\VAN}[3]{#2}
\let\VANthebibliography\thebibliography
\def\thebibliography{\DeclareRobustCommand{\VAN}[3]{##3}\VANthebibliography}

\usepackage{graphicx}	
\usepackage{amsmath}	
\usepackage{subfigure}
\usepackage{color}
\usepackage[dvipsnames]{xcolor}

\newcommand{\Msun}{M$_{\odot}$}
\newcommand{\solar}{\emph{Solar}}
\newcommand{\survey}{\emph{Survey}}

\title[Kinematics of Milky Way Stellar Halo]{The Kinematic Properties of Milky Way Stellar Halo Populations}

\author[J. M. M. Lane et al.]{
James M. M. Lane$^{1}$\thanks{E-mail: lane@astro.utoronto.ca},
Jo Bovy$^{1}$, \&
J. Ted Mackereth$^{1,2,3}$
\\
$^{1}$Department of Astronomy and Astrophysics, University of Toronto, 50 St. George Street, Toronto ON, M5S 3H4, Canada\\
$^{2}$Canadian Institute for Theoretical Astrophysics, University of Toronto, 60 St. George Street, Toronto ON, M5S 3H8, Canada\\
$^{3}$Dunlap Institute for Astronomy and Astrophysics, University of Toronto, 50 St. George Street, Toronto ON, M5S 3H4, Canada
}

\date{Accepted XXX. Received YYY; in original form ZZZ}

\pubyear{2021}

\begin{document}
\label{firstpage}
\pagerange{\pageref{firstpage}--\pageref{lastpage}}
\maketitle

\begin{abstract}

In the \textit{Gaia} era stellar kinematics are extensively used to study Galactic halo stellar populations, to search for halo structures, and to characterize the interface between the halo and hot disc populations. We use distribution function-based models of modern datasets with 6D phase space data to qualitatively describe a variety of kinematic spaces commonly used in the study of the Galactic halo. Furthermore, we quantitatively assess how well each kinematic space can separate radially anisotropic from isotropic halo populations. We find that scaled action space (the ``action diamond'') is superior to other commonly used kinematic spaces at this task. We present a new, easy to implement selection criterion for members of the radially-anisotropic \textit{Gaia}-Enceladus merger remnant. Assuming a 1:1 ratio of \textit{Gaia}-Enceladus stars to more isotropic halo, we find our selection achieves a sample purity of 82 per cent in our models with respect to contamination from the more isotropic halo. We compare this criterion to literature criteria, finding that it produces the highest purity in the resulting samples, at the expense of a modest reduction in completeness. We also show that selection biases that underlie nearly all contemporary spectroscopic datasets can noticeably impact the $E-L_{z}$ distribution of samples in a manner that may be confused for real substructure. We conclude by providing recommendations for how authors should use stellar kinematics in the future to study the Galactic stellar halo. 
\end{abstract}




\section{Introduction}
\label{sec:Introduction}

It is an auspicious time to study the dynamics and structure of our Milky Way galaxy, in no small part thanks to a recent wealth of data obtained through large astronomical surveys. Chief among these is the astrometric space mission \textit{Gaia} \citep{gaia} which, as part of its most recent data release, has provided 5-d kinematic phase space information for nearly 1.5 billion stars, and full 6-d phase space information for a further $\sim7$ million stars \citep{gaiaedr3}. In parallel to this project, massive spectroscopic campaigns such as APOGEE \citep{apogee}, LAMOST \citep{lamost}, Gaia-ESO \citep{gaiaeso}, GALAH \citep{galah}, SEGUE \citep{segue}, RAVE \citep[][for the most recent data release]{ravedr16}, and H3 \citep{h3} have provided accurate radial velocities, chemical abundances, and age estimates for hundreds of thousands of these stars. Combined, these two efforts have yielded a rich, multidimensional data set which has not only enabled novel discoveries, but in concert with powerful statistical techniques will allow us to discern among competing models of the structure, history, and composition of each part of our Milky Way galaxy with exquisite detail. 

The nature of the stellar halo, in particular, has come into much sharper focus in the \textit{Gaia} era. Since \textit{Gaia}'s second data release,  a significant amount of evidence has been amassed that the Milky Way was subject to a handful of distinct, significant mergers early in its life. The most well-studied example to date is the colourfully named ``Sausage'', or ``\textit{Gaia}-Enceladus'' (GE), which is a population of metal-rich halo stars with uncharacteristically large radial motions \citep{helmi18,belokurov18,haywood18}. This overdensity stands out in phase space from the remainder of halo stars, which exhibit more isotropically-distributed motions. Estimates of the mass of this population are in the range of $10^{8}$ to $10^{10}$~\Msun \citep{belokurov18,helmi18,fattahi19,deason19,mackereth19a,mackereth20,kruijssen20,naidu21} which would place its stellar mass among the contemporary big satellites of the Milky Way, such as Sagittarius and the Large Magellanic Cloud \citep[e.g.][]{niederste-ostholt10,vandermarel06}.

GE is only one among an assortment of spatially incoherent halo structures which have either been discovered or extensively studied thanks to \textit{Gaia} kinematics and supporting spectroscopy. These include Thamnos \citep{koppelman19b}, the Sequoia \citep{myeong19}, the Helmi streams \citep{helmi99,koppelman18,koppelman19a}, the Inner Galaxy Structure \citep[IGS, also known as Heracles;][]{horta21}, LMS-1 \citep{yuan20}, as well as numerous structures identified by \citet{naidu20}. Paralleling these results driven by stellar data, examination of the Galactic globular clusters population points to a merger known as ``Kraken'' \citep{kruijssen20} or ``Koala'' \citep{forbes20} which predates the above-mentioned events. While many of these structures require further study to confirm their nature and origin, it is nontheless clear that significant merger events play a major role in shaping the stellar halo we see today.

The stellar halo is not solely composed of \textit{ex-situ} accretion remnants, but also \textit{in-situ} stars formed within the Milky Way. Mounting kinematic and chemical evidence suggests that this \textit{in-situ} component consists mostly of stars from the thicker components of the disc \citep{helmi18,haywood18,dimatteo19}. This idea, while not new \citep[e.g.][]{nissen10}, is put into new perspective given the emerging importance of mergers in the early evolution of the Milky Way. Additionally, an emerging class of halo substructures known as ``Splash'' populations, which may be remnants of a proto-disc disrupted by an infalling satellite, may be evidence of the dynamical link between halo and thick disc \citep{belokurov20,naidu20,lian20}.

While each halo population is often discussed as being ``distinct'' in terms of its spatial, kinematic, and/or chemical properties, there is significant overlap in the properties of many, if not all, of these structures. This presents an issue when attempting to dissect the stellar halo on a granular level, as is the intent with current and upcoming studies. We recognize a gap in the established literature, whereby there does not exist simple kinematic models to contextualize these large range of data. N-body models are commonly used to interpret results, but tailoring the simulations to individual datasets is challenging.

In this work we examine mock samples of stars generated from distribution functions which represent the disc and halo populations commonly studied in the Milky Way. Our goals are twofold: first we seek to provide context for observed kinematics from the current and next generation of surveys. Second we use these models to numerically evaluate the best way to separate radially biased halo populations from more isotropic halo populations, an effective proxy for isolating GE remnant stars.

\section{Model Potential and Distribution Functions}
\label{sec:ModelPotentialandDistributionFunctions}

Here we introduce the distribution functions (DFs) and accompanying gravitational potentials that we use to model the Milky Way disc and halo. First, we present the potential and then we describe the DFs and the techniques used to draw kinematic samples from them. Throughout we assume that: the Sun is located at $(R,z,\phi)$ = $(8.178~\text{kpc}, 0.028~\text{kpc}, 0)$ \citep{gravity19,bennett19}, and the solar motion is $(U,V,W) = (11.1, 12.24, 7.25)$~km~s$^{-1}$ \citep{schoenrich10}. We use a left-handed Galactocentric cylindrical coordinate system where angular momentum is positive in the direction of Galactic rotation. By default we express quantities in cylindrical coordinates and are explicit when we do otherwise. The code we use for this project is available on github\footnote{\url{https://github.com/jamesmlane/mw-dfs}}.

\subsection{Milky Way Potential}
\label{subsec:MilkWayPotential}

To represent the Milky Way we use the \texttt{MWPotential2014} model from \citet{bovy15}. The potential has three components. The disc is represented by a \citet{miyamoto75} profile with mass $M_{d} = 6.8\times10^{10}$~\Msun, a scale length of $3$~kpc, and a scale height of $280$~pc. The bulge is an exponentially-truncated power law density profile with mass $M_{b} = 0.5\times10^{10}$~\Msun, power law index $-1.8$, and exponential scale radius of $1.9$~kpc. The halo is an NFW \citep{navarro97} profile with a virial mass of $8\times10^{11}$~\Msun\ and concentration of 15.3. For a discussion of the derivation of the model parameters which describe \texttt{MWPotential2014}, see section~3.5 of \citet{bovy15}.

\subsection{Disc Distribution Functions}
\label{subsec:DiscDistributionFunctions}

We approximate the disc's intricate structure (e.g., \citealt{bovy12}) as a simple ``thin'' and ``thick'' superposition and to represent both disc components we use the quasi-isothermal DF of \citet{binney11} \citep[see also ][]{binney10}. This DF takes the form
\begin{equation}
    f(J_{r},L_{z},J_{z}) = f_{\sigma_{r}}(J_{r},L_{z})f_{\sigma_{z}}(J_{z})\,,
\end{equation}
where $f_{\sigma_{r}}$ largely controls the dynamics within the plane of the disc, and $f_{\sigma_{z}}$ largely controls the vertical dynamics perpendicular to the disc. Each of these are functions of the actions $J_{r}$, $L_{z}$, and $J_{z}$. $f_{\sigma_{r}}$ is given by
\begin{equation}
    f_{\sigma_{r}} = \frac{\Omega \Sigma}{ \pi \sigma_{r}^{2} \kappa } \bigg \rvert_{\small{R_{c}}} [1 + \tanh(L_{z}/L_{0})] \mathrm{e}^{-\kappa J_{r} / \sigma_{r}^{2}}\,,
\end{equation}
where $\Sigma$ and $\Omega$ are the local surface density and orbital frequency respectively. $R_{c}$ and $\kappa$ are the guiding center radius and epicycle frequency of a given orbit. $L_{0}$ is a cutoff angular momentum. $\sigma_{r}$ is the radial velocity dispersion profile. $\Sigma$ has the exponential form
\begin{equation}
    \Sigma(L_{z}) = \Sigma_{0} \mathrm{e}^{(R_{0}-R_{c})/R_{d}}
\end{equation}
where $\Sigma_{0}$ is an arbitrary normalization, $R_{0}$ is a scale radius (Which we set to the location of the Sun), and $R_{d}$ is the scale length of the disc.
The second component, $f_{\sigma_{z}}$, is given by a pseudo-isothermal DF
\begin{equation}
    f_{\sigma_{z}} =  \frac{\nu_{z}}{2\pi \sigma_{z}^{2}} \mathrm{e}^{-\nu_{z} J_{z}/\sigma_{z}^{2}}\,,
\end{equation}
where $\nu_{z}$ and $J_{z}$ are the vertical epicycle frequency and action and $\sigma_{z}$ is the vertical velocity dispersion profile. The parameters we use for the thin and thick disc DFs are summarized in Table~\ref{table:discDFParameters}. Both the radial and vertical exponential profiles have exponential forms given by\footnote{\citet{binney11} have a multiplicative factor $q$ in these exponentials, but it is unity here.}
\begin{equation}
    \begin{split}
        \nonumber \sigma_{r}(L_{z}) = & \sigma_{r,0} \mathrm{e}^{(R_{0}-R_{c})/R_{d}} \\
        \sigma_{r}(L_{z}) = & \sigma_{z,0} \mathrm{e}^{(R_{0}-R_{c})/R_{d}}\,. \\
    \end{split}
\end{equation}
Here $\sigma_{[r,z],0}$ are characteristic velocity dispersions at $R_{0}$. We use, overall, hotter disc DF parameters than are advocated for by \citet{binney11}, but that are not unlike other measurements have suggested \citep[e.g. ][]{bovy12,mackereth19b}. By choosing disc DF parameters on the hot end of what is normally used we hedge our results towards including disc samples which intrude most on the kinematic space normally assigned to the halo.

\begin{table}
    \begin{center}
        \caption{Properties of quasi-isothermal thin and thick disc DFs}
        \label{table:discDFParameters}
        \begin{tabular}{cll}
            Parameter & Thin disc & Thick disc  \\
            \hline
            $R_{d}$          & 3~kpc               & 2~kpc              \\
            $R_{0}$          & 8.178~kpc           & 8.178~kpc          \\
            $\sigma_{r,0}$   & 40~km~s$^{-1}$      & 60~km~s$^{-1}$     \\
            $\sigma_{z,0}$   & 30~km~s$^{-1}$      & 45~km~s$^{-1}$     \\
            $L_{0}$          & 10~kpc~km~s$^{-1}$  & 10~kpc~km~s$^{-1}$ \\
        \end{tabular}
    \end{center}
\end{table}

To sample from this DF, we use \texttt{galpy}\footnote{\url{https://github.com/jobovy/galpy}} \citep{bovy15} where it is implemented as the class \texttt{galpy.df.quasiisothermaldf}. The phase space samples we generate are described further on in \S~\ref{sec:DFSamples}, but for our purposes we are only interested in sampling velocities from this DF given fixed positions. Velocity samples are obtained using rejection sampling of the DF at each individual radius and height above the midplane. Actions for the DF are calculated using the ``St\"{a}ckel fudge'' method of \citet{binney12}. In this approach our Milky Way potential is locally approximated for each individual sample orbit as a St\"{a}ckel potential using a focal length estimated with equation~(9) of \citet{sanders12}.

\subsection{The Stellar Halo Distribution Function}
\label{subsec:HaloDistributionFunction}

For the stellar halo we use spherical DFs with constant anisotropy. These types of DFs generally take the form
\begin{equation}
    f(\mathcal{E},L) = L^{-2\beta} f_{1}(\mathcal{E})\,.
\end{equation}
Conventionally, here, $\mathcal{E} = \Psi-\frac{1}{2}v^{2}$ is the binding energy, where $\Psi = -\Phi+\Phi_{0}$ is the negative of the gravitational potential offset to be 0 at infinity, and $v$ is the magnitude of the velocity. $L$ is the total 3-dimensional angular momentum. The parameter $\beta$ is the orbital anisotropy, defined as
\begin{equation}
    \beta = 1- \frac{\sigma^{2}_{\theta} + \sigma^{2}_{\phi}}{2\sigma^{2}_{r}}\,,
\end{equation}
where $\sigma_{[\theta,\phi,r]}$ are the spherical velocity dispersions in the polar, azimuthal, and radial directions respectively. $\beta$ ranges from $-\infty$ (completely tangential orbits), over 0 (ergodic, isotropic, or unbiased orbits), and up to 1 (completely radial orbits). As shown in chapter 4.3 of \citet{binney08} the function $f_{1}$ is related to the mass density of the DF\footnote{Note they use number density instead of mass density, but that just amounts to an overall normalization of the DF}, $\rho$ as
\begin{equation}
    \label{eq:AbelIntegral}
    \frac{ 2^{\beta-1/2} }{ 2\pi I_{\beta} } r^{2\beta}\rho = \int_{0}^{\Psi} \mathrm{d}\mathcal{E} \frac{ f_{1}(\mathcal{E}) }{ (\Psi-\mathcal{E})^{\beta-1/2} }\,.
\end{equation}
$I_{\beta}$ is a constant given by
\begin{equation}
    I_{\beta} = \sqrt{\pi}\frac{\Gamma(1-\beta)}{\Gamma(3/2-\beta)}\,,
\end{equation}
where $\Gamma$ is the usual Gamma function.

For $1/2 < \beta < 1$, a DF of this form is an Abel integral equation that can be inverted to yield, often numerically, a solution for $f_{1}(\mathcal{E})$. For lower values of $\beta$, one or more derivatives with respect to $\Psi$ turn this expression into an invertible Abel integral equation \citep{cuddeford91}. Note that while we can use this DF for $\beta > 1/2$, generally it has an increasingly large region where the DF is negative and, thus, unphysical. We set the DF to zero wherever the expression gives a negative number. This causes a slight, but unimportant inconsistency between the DF and the density profile. For $\beta=0.9$ (the highest value we will consider) the profile becomes shallower, with density decreasing by about 10 per cent at 3~kpc being roughly unchanged at the solar circle, and increasing by a few per cent at more distant radii.

For many implementations of this DF where the system is self-gravitating the potential, $\Psi$, and the density profile, $\rho$, are a potential-density pair. This is not necessary, however, and $\rho$ may be independent from $\Psi$. For our DFs $\rho$ is an exponentially truncated power law density profile of the form
\begin{equation}
    \rho(r) \propto r^{-\alpha}\mathrm{e}^{r/r_{c}},
\end{equation}
where $\alpha$ is the inner power law slope and $r_{c}$ is the truncation radius. The overall normalization of the density profile is unimportant because it only determines positions and kinematics of samples in a relative sense. We set $\alpha=3.5$ and $r_{c}=30$~kpc. Our choice of parameters is inspired by \citet{mackereth20}, who find $\alpha=3.49$ and $r_{c}=25$~kpc fit a large sample of halo stars spanning both [Fe/H] and [$\alpha$/Fe]. These fits stand in broad agreement with prior results, both in terms of the power law slope, and the transition radius at which the density profile steepens \citep{deason11,xue15,iorio18}.

The potential in the calculation of these stellar-halo DFs is spherical. Because our adopted \texttt{MWPotential2014} model is non-spherical, we construct a spherically symmetric analog of this potential by calculating its enclosed mass within spherical shells as a function of Galactocentric radius $M(<r)$. The spherical version of the potential is then defined as $\Phi_{MW} = -\frac{GM(<r)}{r}$. While possible, it is difficult to create anisotropic DFs for spherical systems, such as the stellar halo, with a given density and anisotropy embedded within axisymmetric, non-spherical potentials \citep{posti15,williams15}. Our approach, while approximate, is simple to implement and gives us greater control over the kinematic and spatial properties of our halo DFs. The act of sphericalizing the potential should have a minimal impact on our results, which focuses on kinematic properties which are sensitive to the total enclosed mass of the potential rather than increased mass near the disc plane. 

We sample from DFs of this type using \texttt{galpy}, where they are implemented as the class \texttt{galpy.df.constantbetadf}. More details on our implementation of these spherical DFs and how we specifically sample from them can be found in Appendix~\ref{dfappendix}.

\section{DF Samples}
\label{sec:DFSamples}

In this section, we present the samples used to explore the observable kinematics of Milky Way stellar populations, and describe how we calculate their kinematics. We first outline the overall properties of our samples and then describe how we generate the two specific sets of samples that we use throughout the rest of the paper. Finally, we present the kinematics of each of these samples.

Broadly speaking our samples consist of phase space positions drawn from a combination of the distribution functions outlined in \S~\ref{subsec:DiscDistributionFunctions} and \ref{subsec:HaloDistributionFunction}, representing the disc (approximated as a thin plus thick disc) and stellar halo. We do not consider the stellar bulge for two reasons. First, the bulge does not contribute stellar density near the solar neighbourhood in any meaningful capacity, and therefore bulge stars are reasonably easy to remove from real stellar samples by cutting stars in angular proximity to the Galactic centre. This means one can be confident that the bulge does not represent a substantial source of contamination when studying halo populations, in contrast to the stellar disc. Second, the complexity of bulge kinematics due, for example, to the presence of the rotating Galactic bar means that it is challenging to realistically model bulge populations using DFs.

\subsection{The \solar\ Samples}

 The first set of samples, which we refer to as \solar\ are located at the Sun's position: ($R,z$) = (8.178~kpc, 20.8~pc) \citep{gravity19,bennett19}. This set is designed to be instructive, and so having all samples share the same position simplifies the kinematics. Yet, because spatial gradients in the DF occur over multiple-kpc scales, this sample is also somewhat realistic in that it represents stellar samples confined to the $\approx 1$~kpc region around the Sun, a distance over which the halo DF varies little. This includes existing samples such as the \textit{Gaia} radial velocity sample \citep[see section~3.3 in  ][ for the most recent data]{gaiaedr3} or GALAH \citep{galah,galahdr3}. The \solar\ set of samples consists of six independent samples of stars, which differ by the value of $\beta$ used for the stellar halo DF.

Each of the \solar\ samples consists of $10^{5}$ thin disc orbits and $1.2\times10^{4}$ thick disc orbits, sampled according to the procedure outlined in \S~\ref{subsec:DiscDistributionFunctions}. These numbers are chosen such that the ratio of thick to thin disc stars is approximately 1:8.5, which reflects the ratio of thick to thin disc stars expected at the solar circle \citep[e.g. see the potential of][]{mcmillan17}. Each of the samples also contains $10^{5}$ orbits sampled from our spherical halo DF with a different value of $\beta$. We are not concerned with the relative numbers of halo and disc stars, because we will differentiate between the two populations when studying their properties. The first five samples each use a different value of $\beta$ from among $[0,-0.5,0.5,0.7,0.9]$. The sixth sample has a composite stellar halo, also $10^{5}$ orbits total, consisting of a 50:50 mixture of orbits from DFs with $\beta=0.5$ and 0.9 respectively. We refer to this final sample as our fiducial model. 

The ratio of the component DFs in this model follows from the recent Milky Way stellar halo mass measurements of \citet{mackereth20}, who find that the ratio of mass in low and high eccentricity stars is approximately 50:50. These low and high eccentricity populations are interpreted as a traditional metal poor stellar halo with $\beta \approx 0.2-0.4$, and the more metal rich \textit{Gaia}-Enceladus remnant which has been shown to be characterized by large anisotropies of $\beta$ close to 0.9 \citep{belokurov18,fattahi19,iorio21}. This model is broadly consistent with other results that have suggested that these two halo components exist in roughly equal proportion near the solar neighbourhood \citep{belokurov18,lancaster19,iorio21}. By using $\beta=0.5$ to represent the more isotropic component of the halo we err on the side of slightly larger $\beta$, but this allows us to be confident we do not underestimate the degree of contamination from this component when analyzing their kinematics. This is similar to the reasoning behind our choice of slightly hotter disc DF parameters discussed in \S~\ref{subsec:DiscDistributionFunctions}.

Throughout the paper we will refer to the two constituent components of our fiducial model as ``low-$\beta$'' and ``high-$\beta$'' for the $\beta=0.5$ and $\beta=0.9$ halos respectively. While our fiducial model is inspired by literature results which employ metallicity and eccentricity as discriminators for the halo populations, we emphasize that we define our halo components using $\beta$, and therefore they may not be perfectly faithful representations of a halo defined in another manner. 

\subsection{The \survey\ Samples}
\label{subsec:TheSurveySamples}

The second set of samples has physical positions which are matched to the locations of stars in the APOGEE DR16 \citep{apogeedr16} statistical sample, rather than occupying a single location. The statistical sample is, briefly, the subset of APOGEE stars which were selected randomly based on the 2MASS photometric catalogue according to the procedures outlined in \citet[][APOGEE-1]{zasowski13} and \citet[][APOGEE-2]{zasowski17}. We refer to these samples with positions matched to APOGEE data as the \survey\ samples. These samples more realistically represent modern and upcoming datasets where \textit{Gaia} data is augmented by spectroscopy, which allows for the calculation of accurate spectro-photometric parallaxes \citep[e.g.,][]{leung19b} and radial velocities for a small number of stars (compared to the broader \textit{Gaia} dataset) out to large ($\sim 20$~kpc) distances. Examples of these surveys include the contemporary APOGEE \cite{apogeedr16} and H3 \citep{h3} surveys, as well as the future WEAVE \citep{weave}, 4MOST \citep{4most}, DESI \citep{desi} and SDSS-V \citep{sdss5} projects.

We use this APOGEE sample to select an approximate population of halo, thin and thick disc stars, whose positions and distances are used to generate samples in our model. To split the APOGEE sample into halo and disc populations we apply the following selection procedure. We start with the on-sky locations, [Fe/H] and [Mg/Fe] abundances of stars in the APOGEE DR16 statistical sample. We supplement this data with accurate spectro-photometric distances derived using the Bayesian neural network framework \texttt{astroNN} \citep{leung19a,leung19b} (note these are based on \textit{Gaia} DR2 \citep{gaiadr2}). Our goal is to use this information to obtain a rough set of positions corresponding to three populations we seek to study: the halo, thin disc, and thick disc.

\begin{figure}
	\centering
	\includegraphics[width=\columnwidth]{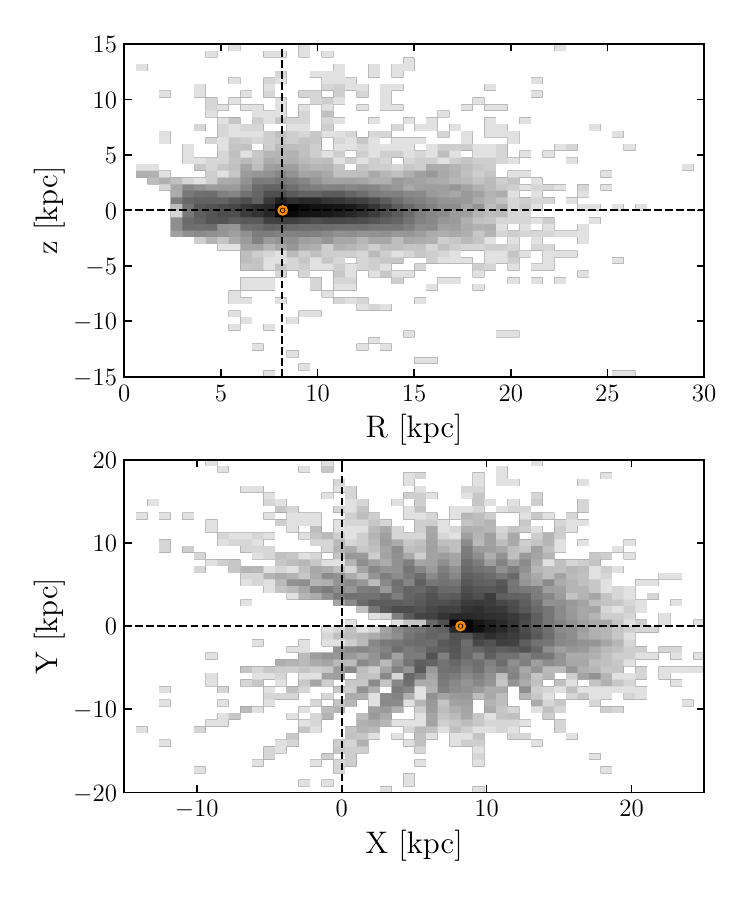}
	\caption{The APOGEE DR16 sample shown in Galactocentric coordinates. The top panel shows Galactocentric cylindrical radius and height above the disc, the bottom panel shows Galactocentric X and Y, where X is along the Sun-Galactic Centre line, and Y is positive in the direction of Galactic rotation. In both panels the Sun is shown at (X,Y,Z) = (8.178~kpc, 0, 0.028~kpc).}
	\label{fig:APOGEELocations}
\end{figure}

We perform a number of cuts on the APOGEE data to ensure both quality and that the underlying stellar populations correctly reflect our model DFs. The cuts are as follows. The $\log g$ uncertainty is less than 0.1, which removes many dwarfs with poorly constrained properties. The relative distance uncertainty, $\sigma_{d}/d$, is less than 0.2. We remove fields near the Galactic bulge, which we define as those with $-20^{\circ} < \ell < 20^{\circ}$ and $|b| < 20^{\circ}$. We also remove any fields which contain distinct groups of stars from a globular cluster. We first identify which APOGEE fields contain the on-sky location of a known globular cluster from the catalogue of \citet[][ The December 2010 version]{harris96}. We then examine the distribution of [Fe/H] abundances and radial velocities for the stars in the field, in particular those within two tidal radii of the globular cluster location. Fields where there is an obvious enhancement of stars with the appropriate [Fe/H] and radial velocities near the on-sky location of the globular cluster are removed from consideration. Figure~\ref{fig:APOGEELocations} shows the spatial distribution of the final selection of APOGEE stars. Our sample spans a wide range of the Milky Way, nearly 15~kpc in each direction from the location of the Sun.

\begin{figure}
	\centering
	\includegraphics[width=\columnwidth]{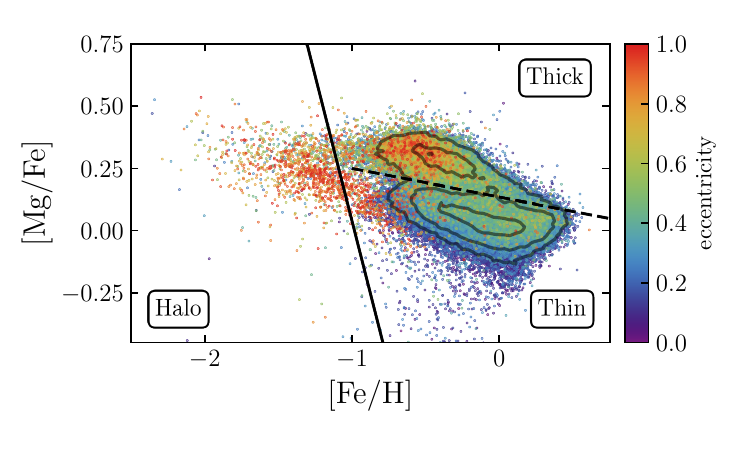}
	\caption{[Mg/Fe] versus [Fe/H] abundances for the APOGEE DR16 sample. Each star is color-coded by its eccentricity, with higher eccentricity stars appearing above lower eccentricity stars. The solid line shows the separation between halo and disc stars. The dashed shows the separation between thin and thick disc stars. The grey contours show density in the more populated thin and thick disc sequences.}
	\label{fig:APOGEEAbundances}
\end{figure}

We separate the APOGEE stars into three categories: halo, thin disc, and thick disc using their [Fe/H] and [Mg/Fe] abundances as well as their velocities. We stress that the division into a thin and thick disc is purely an approximation to the more complex structure of the disc, but one that is good enough for our main purpose of investigating the stellar halo. Figure~\ref{fig:APOGEEAbundances} shows these abundances for our final APOGEE sample, as well as the boundaries we adopt to separate each stellar population. We choose these boundaries based on the number density of APOGEE stars in the abundance space, as well as their eccentricities (which color-codes the points in the top panel of Figure~\ref{fig:APOGEEAbundances}). The thin-thick disc boundary follows the upper edge of the low-[Mg/Fe] disc locus, while the disc-halo boundary follows the low-[Fe/H] edge of the disc locus and approximately matches the transition from low to high eccentricity. We also add to the halo sample any stars lying in the thin or thick disc chemical regions which have net velocities with respect to their local circular velocity frame which are greater than the magnitude of their local circular velocity: $\lvert \vec{v} - \vec{v}_\mathrm{circ}(R) \rvert > \lvert \vec{v}_\mathrm{circ}(R) \rvert $. We find that this efficiently attributes many higher metallicity stars with genuine halo kinematics to the halo sample. We examine the energy and angular momentum plane, as well as the tangential and radial velocity plane for each of these selections to ensure that they reflect what we expect from the kinematics of the respective populations. Overall, however, the accuracy of this separation is not a major concern, because this only sets the positions based on which velocities will be sampled from the individual DFs. 

The final separated samples consist of 124,879 thin disc stars, 23,543 thick disc stars, and 2,198 halo stars. Like the \solar\ sample, the \survey\ samples consist of six individual sets of orbits which differ by the value of $\beta$ used for the halo DF. Again, the first five samples each use a different value of $\beta$ from among $[0,-0.5,0.5,0.7,0.9]$, while the sixth sample has a composite stellar halo consisting of a 50:50 mixture of orbits from DFs with $\beta=0.5$ and $0.9$ respectively.

To add a degree of realism to the \survey\ samples we perturb all (thin disc, thick disc, and halo) of the velocities using the uncertainties of the \textit{Gaia} and APOGEE DR16 data to which these samples are matched. We do not perturb sample positions, since they are matched to data which are naturally subject to positional uncertainty, and therefore to perturb the positions would be to duplicate the effect. Velocities, on the other hand are drawn from the DFs, and so are not subject to any uncertainties. We first convert the sample velocities into observer coordinates: ($\mu_\mathrm{RA}$, $\mu_\mathrm{Dec}$, $v_\mathrm{los}$), and then add to each of the vectors an amount sampled from a multivariate Gaussian with zero mean and covariance matrix composed of the uncertainties and correlations drawn from the \textit{Gaia} DR2 and APOGEE data. We assert that there are no correlations between APOGEE line-of-sight velocities and any of the \textit{Gaia} data.

\subsection{Sample Kinematics}
\label{subsec:SampleKinematics}

For both the \solar\ and \survey\ set of samples we calculate kinematic quantities as well as actions using \texttt{galpy} \citep{bovy15}. Actions are calculated using the ``St\"{a}ckel fudge'' method of \citet{binney12}. In this approach our Milky Way potential is locally approximated for each individual sample orbit as a St\"{a}ckel potential using a focal length estimated with equation~(9) of \citet{sanders12}. Eccentricities are similary calculated using a variation of the ``St\"{a}ckel fudge'' method as described in \citet{mackereth18c}. These actions and eccentricities are calculated in the axisymmetric \texttt{MWPotential2014} potential. 

Potentials like that of the Milky Way are well-represented by St\"{a}ckel potentials \citep{dejonghe88}, and the `St\"{a}ckel fudge'' approximation gives more accurate action estimates for general orbits when compared with the spherical and adiabatic approximations, which are only good approximations when the potential is close to spherical, and the orbit is confined near the symmetry plane of an axisymmetric potential respectively. As shown by \citep{bovy15} the variation in St\"{a}ckel method action estimates throughout an orbit is less than about 2~per cent for a disc-like orbit with reasonable radial and vertical extent. \citet{mackereth18c} find that the St\"{a}ckel approximation performs just as well as direct orbit integration for essentially all halo orbits and most disc-plane orbits which do not venture near the center of the galaxy (where resonances tend to complicate the orbit structure). We test the constancy of this approximation by calculating focal lengths and actions for a few randomly selected APOGEE halo stars at multiple points along their orbits. We find the typical median absolute deviation to be 7~kpc~km~s$^{-1}$ for the radial actions and 3~kpc~km~s$^{-1}$ for the vertical action ($L_{z}$ is obviously invariant along the orbit). The typical median absolute deviations expressed relatively are 2 per cent for the radial actions and 1 per cent for the vertical action, which is similar to the findings of \citet{bovy15}.

We consider six kinematic spaces which are commonly used throughout the literature for both structure identification and classification, particularly of halo populations. The first two use only velocity information, while the latter four rely on conserved quantities calculated using the underlying potential. For each space we provide a few recent example studies which have made use of it, but the list is by no means exhaustive. The first is the $v_{R}-v_{T}$ plane \citep[e.g. ][]{belokurov18, fattahi19,lancaster19,mackereth19a,belokurov20, feuillet20}, which is useful since disc and halo populations occupy distinct loci defined by their mean $v_{T}$, and disc populations have much lower tangential and radial velocity dispersions than halo populations. Additionally the degree of radial bias in a halo population is easy to gauge by the radial extent of its velocity ellipsoid. The second space is the Toomre diagram \citep[e.g. ][]{hawkins15,helmi18,koppelman19b,feuillet20,cordoni20} which plots the quadrature $\sqrt{v_R^2+v_z^2}$ of the $v_{R}$ and $v_{z}$ velocities (also known as $v_\mathrm{perp}$) as a function of $v_{T}$. Here we use cylindrical velocities, yet the Toomre diagram is also often expressed using $UVW$ velocities. In samples that span a large range in Galactic azimuth, the $v_R$ and $v_T$ velocities project onto $U$ and $V$ in a manner that renders the diagram less useful. The advantage of the Toomre diagram is that it factors in $v_{z}$ in addition to $v_{R}$, further isolating halo populations thanks to the relatively cold vertical velocity kinematics of the thin and thick disc populations.

The next three spaces all use angular momentum about the disc rotation axis, $L_{z}$, as the abscissa. $L_{z}$ is an effective coordinate since disc stars tend to have characteristically large, positive values. Additionally, because it is one of three actions, it is conserved under adiabatic conditions. The third space is the $E-L_{z}$ plane \citep[e.g. ][]{helmi18,koppelman19b, naidu20,feuillet20,cordoni20,horta21}, which is perhaps the most widely used kinematic space for structure identification in the \textit{Gaia} era. This is due to the relative ease with which these parameters are calculated, their approximate conservation through cosmic time, the fact that the disc occupies a compact locus at high $L_{z}$ and moderate $E$, and that the energy encodes both radius and velocity information. We plot all energies offset by the potential of \texttt{MWPotential2014} at infinity, $\Phi_{0}$, so that $E<0$ are bound orbits. The fourth space is the root of the radial action, $\sqrt{J_{R}}$, as a function of $L_{z}$ \citep[e.g.][]{trick19,feuillet20,matsuno21,horta21}. By using the radial action this space can very efficiently separate halo stars on tangential versus radial orbits. This is because the action labels the entire orbit while velocities, for example, will change throughout an orbit. We use the square root of the radial action to compactify the vertical axis for efficient visualization. The fifth space is eccentricity ($e$) as a function of $L_{z}$ \citep[e.g.][]{cordoni20}. While this specific two-dimensional space is not commonly used, eccentricity as a single parameter is often used as a metric to separate GE from the remainder of the stellar halo since it reflects the overall radial extent of an orbit \citep[e.g.][]{belokurov18,mackereth20,naidu20}.

The final space is the ``action diamond'' \citep[e.g. ][]{vasiliev19,myeong19,monty20,cordoni20,naidu20} which in our implementation plots $J_{z}-J_{R}$ as a function of $L_{z}$, the azimuthal action. But both quantities are scaled by the total action defined as $J_\mathrm{tot}=\lvert J_{R} \rvert + \lvert J_{z} \rvert + \lvert L_{z} \rvert$, which causes the appearance of a characteristic diamond-shaped boundary (Note that if the normalizing total action is $J_\mathrm{tot}= \sqrt{ J_{R}^{2} + J_{z}^{2} + L_{z}^{2}}$ the boundary is circular, e.g., \citealt{naidu20}). One advantage of this space is that the quantities are scaled by their total action, which means that disc populations whose action budget is dominated by $L_{z}$ tend to cluster tightly in the right corner of the bounding diamond. For the same reason halo sub-populations which arise from, for example a single merger event, are likely to cluster at a single location defined by the actions of the progenitor. The downside of this space is that since all quantities are normalized by the total action, populations which differ by an overall magnitude in their actions, such as eccentric disc and halo populations, may appear indistinct. To combat this effect it is often adviseable to use this space only with a sample confined to a narrow range of radii or $L_{z}$.

It is natural to consider whether higher dimensional spaces would be more appropriate for this type of study. While we have mentioned above many unique quantities that describe stellar orbits, they are not all completely independent. We believe that there are three broad classes of quantities here that we consider here. First are quantities that describe tangential motion, which includes $v_{T}$ and $L_{z}$ (whether it is normalized by $J_\mathrm{tot}$ or not. The second are quantities which describe the magnitude of radial motion, which include $v_{R}$, $v_\mathrm{perp}$, energy, and $J_{R}$. The third are scaled quantities which describe the radiality of an orbit, which include the normalized quantity $(J_{z}-J_{R})/J_\mathrm{tot}$ and eccentricity. Within each of these categories is a modest to large amount of degeneracy. For example a star with high radial velocity likely also has a large $v_\mathrm{perp}$, energy, and $J_{R}$. On the other hand, a star with large radial velocity may not necessarily be on a particularly radial orbit if its tangential velocity is also large. This suggests that the largest dimensional space which would be useful in a coherent sense would be three dimensional. We will explore the possibility of using higher dimensional spaces to select high purity samples of stars from high-$\beta$ populations in \S~\ref{subsec:SeparatingHighBetaPopulations}. For now we will continue to work with two dimensional spaces for ease of visualization.

\begin{figure*}
	\centering
	\includegraphics[width=\textwidth]{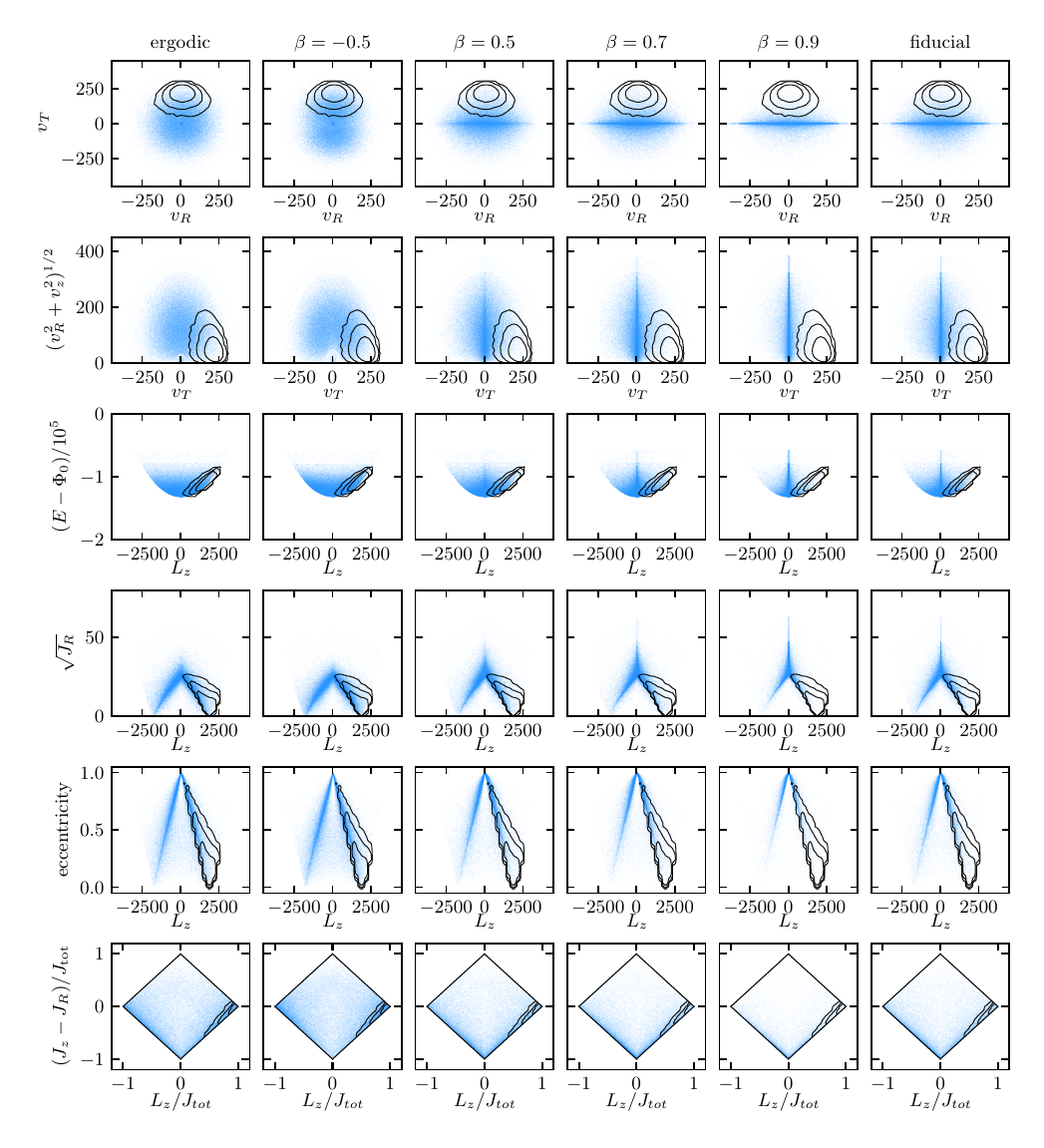}
	\caption{Kinematics of the six \solar\ samples. Each of the columns shows one of the subsets of the \solar\ sample, which vary by the $\beta$ of the halo component (labelled). The blue points show the halo samples while the black contours show the disc samples, combining both thin and thick. \textit{Top row:} Radial and tangential velocities; \textit{second from top:} Toomre diagram; \textit{third from top:} energy and $L_{z}$; \textit{third from bottom:} $\sqrt{J_{R}}$ and $L_{z}$; \textit{second from bottom:} eccentricity and $L_{z}$; \textit{bottom row:} action diamond. The units (unlabelled) are as follows: velocities are in km~s$^{-1}$, actions and angular momenta are in kpc~km~s$^{-1}$, and energies are in km$^{2}$~s$^{-2}$.} \label{fig:SolarKinematics}
\end{figure*}

\begin{figure*}
	\centering
	\includegraphics[width=\textwidth]{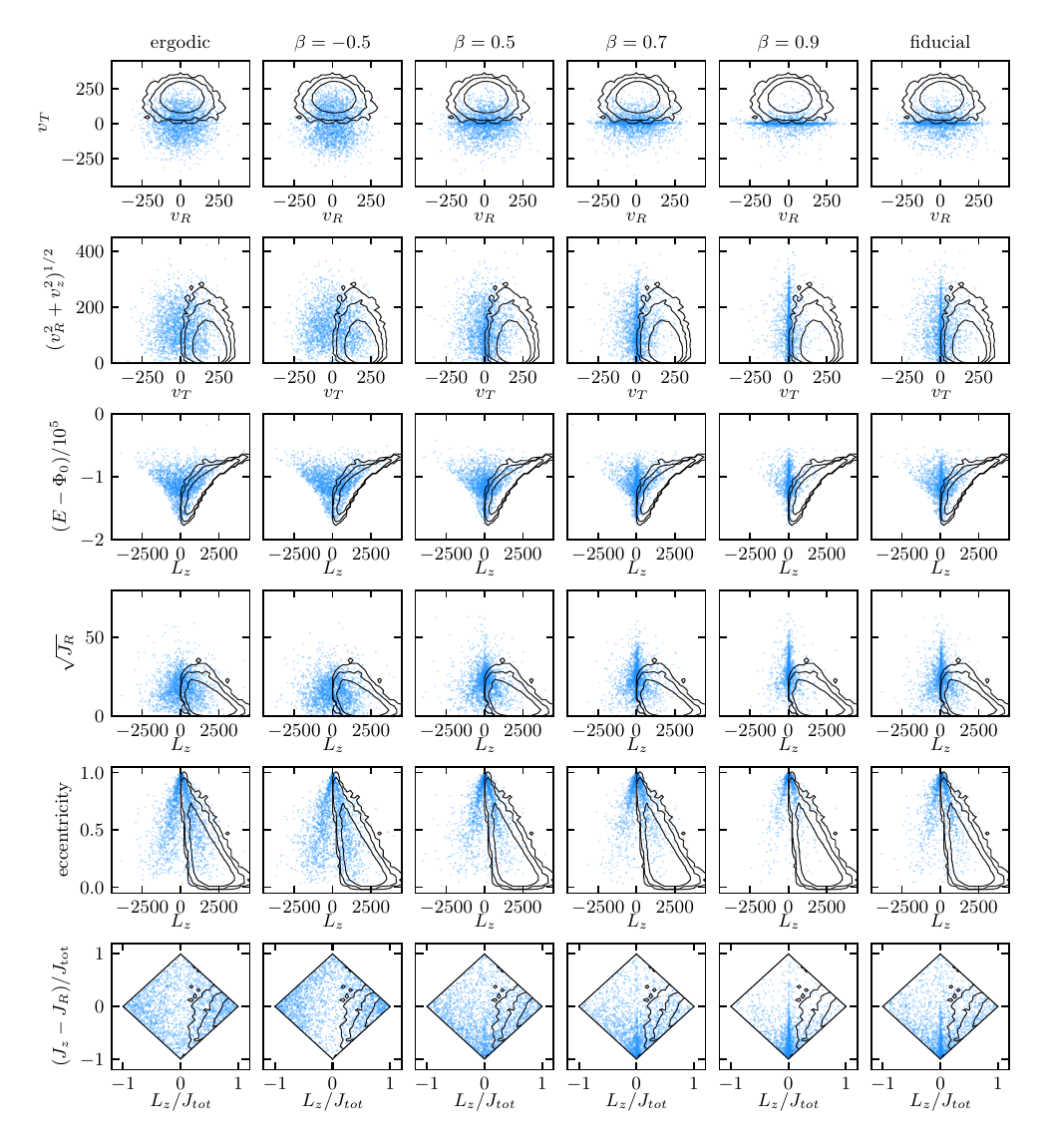}
	\caption{Same as Figure~\ref{fig:SolarKinematics} but for the six \survey\ samples.}
	\label{fig:SurveyKinematics}
\end{figure*}

Figure~\ref{fig:SolarKinematics} shows these six kinematics spaces for the \solar\ samples. The disc samples, both thin and thick together, are shown as the black contours, while the halo samples are shown as blue points. Each of the columns shows one of the six different halo models. Figure~\ref{fig:SurveyKinematics} is the same as Figure~\ref{fig:SolarKinematics}, but shows the kinematics for the \survey\ samples instead. The disc sample contours are constructed by binning the data with 40 bins across each dimension in each space. The contour levels are then placed at 1, 10, and 100 samples per bin. 

\section{Exploring the Kinematic Properties of our Samples}

In this section, we examine in detail each of the kinematic spaces we presented in the previous section. We first provide a qualitative overview of the kinematic spaces, especially with regards to how kinematics change with variations in the value of $\beta$ of the stellar halo DF. 

\subsection{A Qualitative Assessment of the Kinematic Spaces}

\subsubsection{The Solar Samples}

We first take a closer look at the \solar\ sample kinematics from Figure~\ref{fig:SolarKinematics}. Starting with the $v_{R}-v_{T}$ velocities, we can immediately see a strong degree of evolution in the morphology of the halo samples as $\beta$ changes. Specifically, as $\beta$ increases the dispersion along the $v_{R}$ axis increases and the dispersion along the $v_{T}$ axis decreases, broadly in accordance with the definition of $\beta$. 

One subtle feature is that as $\beta$ increases the density of samples along the $v_{T}=0$ line forms a broad plateau around $v_{R}=0$, rather than the distinct density peak that can be seen when $\beta=0$ for example. This behaviour is the result of a tendancy towards bimodality in the $v_{R}$ direction for high-$\beta$ samples, since stars are on more radial orbits and therefore preferentially have non-zero radial velocities (with equal preference to positive and negative velocities). Examining the fiducial model highlights why it is important to keep this fact in mind. When low and high-$\beta$ populations are shown together, it may appear to be a superposition of two centrally concentrated ellipsoids, yet in reality the lower-$\beta$ population provides the peak at $(v_{R},v_{T}) = (0,0)$ while the radially biased population tends to more evenly populate the low and high radial velocity region. For much higher $\beta$ or when the density profile allows for apocenters at much larger Galactocentric radii, one will also expect to observe a genuine bimodality in the radial velocity direction, with a distinct lack of stars with zero velocity. 

This is noteworthy when modelling the local stellar halo velocity field using, for example, a Gaussian mixture model. One would expect a two-component model to fit the radially anisotropic halo better than a single component. This assessment is supported by the results of \citet{belokurov18} and \citet{fattahi19}, who find that using two Gaussian components for the local halo (i.e. one for the metal poor, more isotropic halo, and one for the radially anisotropic halo) produces best-fits which have systematically negative model residuals near $v_{r}=0$ (note they use spherical polar coordinates), and positive model residuals towards larger radial velocities. This indicates that the single component fit to the radially anisotropic population expects more stars near $v_{r}=0$ and fewer at large radial velocities, in accordance with our expectations. Other studies which have used symmetric, bimodal Gaussians have found good fits to data \citep{lancaster19,necib19}.

The Toomre diagram shows evolution with increasing $\beta$ in a manner similar to the $v_{R}-v_{T}$ plane. As the radial anisotropy of the halo population increases the distribution elongates along the $v_\mathrm{perp}$ axis, and narrows along the $v_{T}$ axis. The inclusion of $v_{z}$ information in addition to $v_{R}$ in the Toomre diagram does tend to separate the halo samples from the disc samples somewhat more than in the $v_{R}-v_{T}$ plane. But the effect is not as pronounced as might be expected, likely because many disc samples have roughly comparable vertical and radial velocities, and so their extent along the $v_\mathrm{perp}$ axis also increases. In both of these first two kinematic spaces, the disc occupies similar loci at $v_{T} \sim 200$~km~s$^{-1}$ with zero net radial or vertical velocity. It is important to note, however, that even in this population of disc samples located at the solar position there are a marginal yet noticable number belonging to the thick disc with near-0 $v_{T}$. In the Toomre diagram even though the halo samples are more distinct, the maximum perpendicular velocities of the hottest thick disc samples approaches 200~km~s$^{-1}$. 

The $E-L_{z}$ plane in the third row of Figure~\ref{fig:SolarKinematics} exhibits the now famous central enhancement of samples with $L_{z} \approx 0$ for large $\beta$. It is this feature which is most characteristic of the GE population as first shown by \citet{helmi18}. One of the reasons this space has been studied so frequently in the \textit{Gaia} era is that both $E$ and $L_{z}$ are approximately conserved integrals of motion, and therefore the appearance of stellar (sub)structures in this space often point to a common ancestry for the constituent stars. Another key feature of this space which we will explore in more depth later is the distinct parabolic shape of the overall distribution. This boundary is set by the fact that for a given energy and radius the magnitude of the velocity is fixed. This results in a bounding value of $L_{z}$ which occurs when the velocity is wholly oriented in the direction of $v_{T}$. The equation of the bounding curve is: $E = L_{z}^{2}/(2R^{2}) + \Phi(R)$. Because the samples in the \solar\ set are all at a fixed radius the boundary is obvious in Figure~\ref{fig:SolarKinematics}. There is also a clear gradient in energy, with the density of samples decreasing as the energy increases. This is intrinsic to the energy-dependence of spherical DFs \citep[see chapter 4.3 in ][]{binney08} and is therefore observed even at a fixed radius. As the radius changes, the energy of the apex of the bounding parabola also changes. This emphasizes that the relationship between radius and energy is complicated, a feature which we will discuss more later.

The radial action, $J_{R}$, is perhaps the most natural variable for isolating stars from populations with high radial anisotropy, because it labels the entire underlying orbit rather than being sensitive to the phase of the orbit such as velocities are. We see in Figure~\ref{fig:SolarKinematics} that there is an approximately inverse relationship between the magnitude of $L_{z}$ and the radial action. This occurs because as angular momentum increases at a fixed position, less of the velocity budget is necessarily contained in radial motions. The relationship is not perfect, however, for two reasons. First, this kinematic space does not account for vertical motions, and second there is a range of energies in the \solar\ samples as we have just discussed. As $\beta$ increases a distinct cusp forms around $L_{z}=0$ spanning a range of $\sqrt{J_{R}}$. The fact that this cusp has a distinct minima (around $\sqrt{J_{R}}\approx25$ $\mathrm{kpc}^{1/2}~\mathrm{km}^{1/2}~\mathrm{s}^{-1/2}$ in Figure~\ref{fig:SolarKinematics}) reflects the fact that at the solar circle, the only orbits that exist with $L_{z}=0$ must either have a certain degree of radial motion, or otherwise must be on a circular orbit exactly over the Galactic pole. The cusp spans a range of $\sqrt{J_{R}}$ for the same reason that the inverse relationship between $\sqrt{J_{R}}$ and $L_{z}$ is imperfect at non-zero $L_{z}$: there is a range of energies in the samples and vertical motions are not accounted for.

The relationship between eccentricity and $L_{z}$ is similar in spirit to the relationship between $\sqrt{J_{R}}$ and $L_{z}$. As $L_{z}$ increases at a fixed location, eccentricity must necessarily decrease because orbits become less radial. This lends a characteristic v-shape to the samples in this space. Unlike $\sqrt{J_{R}}$, eccentricity is a ``scaled'' parameter, it depends only on the relative values of the orbit's pericenter and apocenter. Therefore, we see that disc samples appear less distinct from halo samples than in other kinematic spaces. From an intuitive perspective, a halo star with pericenter at the solar circle could have the same eccentricity as a disc star with apocenter at the solar circle, yet these two stars would have very distinct radial actions. A second curiosity of eccentricity is the fact that as orbits become more radial, their eccentricity approaches 1 asymptotically slowly. This means that as $\beta$ increases samples bunch up close to $L_{z}=0$ and $e=1$. This is markedly different behaviour than most other kinematic spaces we have presented so far, where the kinematic properties of radially-biased samples have a distinct magnitude. Here, the only distinct property of the radially biased population is its sheer density at high eccentricity. We will discuss how this impacts the use of eccentricity as a parameter to select high-$\beta$ populations later.

The action diamond is perhaps the least used among the kinematic spaces shown here, with a similar implementation first being used by \citet{vasiliev19} \citep[to our knowledge, although they recognize that the space is analogous to visualisations of three-dimensional action spaces presented by][]{binney08}. Its most common use thus far has been to study globular cluster kinematics and their potential association with larger halo structures \citep{vasiliev19,myeong19}. But it still offers useful information for the types of aggregate stellar populations we study here. Understanding the action diamond is intuitive. The right and left points of the diamond ($\lvert L_{z} \rvert =J_\mathrm{tot}$) are in-plane prograde and in-plane retrograde orbits respectively. The top point ($J_{z}=J_\mathrm{tot}$) is a polar orbit, and the bottom point ($J_{R}=J_\mathrm{tot}$) is a radial orbit. The bottom-right and bottom-left edges ($J_\mathrm{tot}$ wholly contained in $L_{z}$ and $J_{R}$) are prograde and retrograde in-plane orbits. The top-right and top-left edges ($J_\mathrm{tot}$ wholly contained in $L_{z}$ and $J_{z}$) are prograde and retrograde out-of-plane orbits.

One of the most noticeable features of the action diamond is that both disc and halo populations tend to be biased towards the bottom half of the diamond, which reflects the fact that both populations tend to have modestly hotter radial versus vertical kinematics. Indeed, we see what is perhaps the most subtle transformation in morphology from the low to high-$\beta$ halo population, where the samples appear to concentrate more towards the bottom point of the diamond rather than the left and right points. This results from the dominance of the radial action over the vertical action in the high-$\beta$ sample. The disc also hugs the bottom-right edge of the diamond, but the action budget is dominated by $L_{z}$, meaning it is mainly confined to the right point of the diamond. Similarly to eccentricity this is also a ``scaled'' space, so two orbits with different absolute kinematic properties may occupy the same point. So, while the strength of the action diamond lies in the fact that distinct types of orbits occupy different loci in the space, the weakness is that confusion between two populations that may have different absolute kinematics, but similar types of orbits is likely. This will become more evident when examining the \survey\ sample set.

\subsubsection{The \textit{Survey} Samples}

The \survey\ samples differ from the \solar\ samples in three key respects with regards to the six kinematic spaces that we consider. First, the sample positions span a range of Galactocentric radii, and so in some sense it is useful to consider the kinematic profiles as a superposition of many of the individual samples seen in Figure~\ref{fig:SolarKinematics}, but determined at different radii. Second, the samples are perturbed by \textit{Gaia} and APOGEE errors. Third, and perhaps most important, is that the positions underlying the samples are subject to the complicated APOGEE DR16 selection function. It is through these three lenses that we will assess the kinematics of the \survey\ sample set in Figure~\ref{fig:SurveyKinematics}.

First, one obvious change to all of the kinematic spaces is the increased range over which the disc populations are observed, while in the \solar\ samples the disc populations tended to be confined to a narrow region. This is due mostly to the first two factors mentioned above: the increased radial range over which the DFs are sampled and the addition of realistic uncertainties. Furthermore the discrepancy in the number of disc samples versus halo samples (roughly 66:1) exacerbates the degree to which the disc populations extend into the stellar halo kinematic space with meaningful density.

In the $v_{R}-v_{T}$ plane, the overall morphology of the halo samples stays the same, with bimodality along the $v_{R}$ axis increasing as $\beta$ increases. The impact of the realism steps on the data is clear, particularly in how the \survey\ samples do not cluster so tightly on the $v_{T}=0$ line, especially in the samples with higher $\beta$, as the \solar\ samples do. Overall our fiducial model looks realistic in its representation of the nearby Milky Way $v_{R}-v_{T}$ plane \citep[c.f. figures 2 and 1 in ][ respectively, although note they use spherical coordinates]{belokurov18,fattahi19}, a testament both to the validity of our models and the steps we take to add realism to our DF samples. The Toomre diagram changes in much the same way as the $v_{R}-v_{T}$ plane: the overall morphology is the same, but the addition of uncertainties has softened the cusp in the halo samples along the $v_{T}=0$ line. In both these planes, the disc population now extends down to near $v_{T}=0$, which reflects the addition of samples from the inner galaxy where the angular momentum is lower and the radial velocity dispersion is higher. The extent of the disc in the $v_{R}$ and $v_\mathrm{perp}$ dimensions does not increase as drastically, because only the increased radial and vertical velocity dispersions in the inner Galaxy serve to modify its extent.

\begin{figure}
    \centering
    \includegraphics[width=\columnwidth]{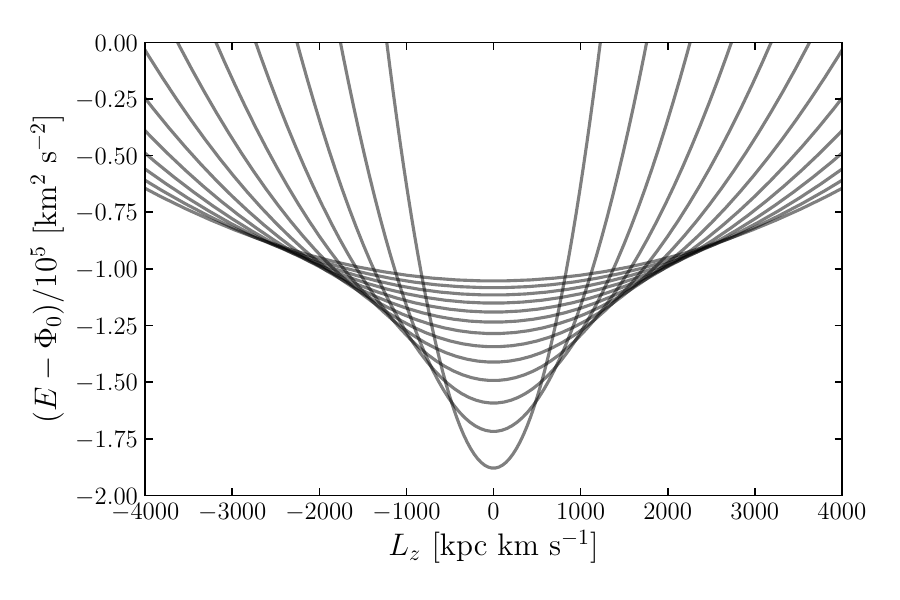}
    \caption{The $E-L_{z}$ plane shown with bounding parabolas corresponding to Galactocentric radii 2 (lowest curve) to 15 (highest curve) kpc. The curves are calculated in-plane ($z=0$), in \texttt{MWPotential2014}.}
    \label{fig:ELzBoundaries}
\end{figure}

From a qualitative standpoint, the $E-L_{z}$ plane changes in much the same way as the velocity planes. The cuspy overdensity of the high-$\beta$ populations is softened, and the hottest thick disc stars begin to impinge on this characteristic region, yet leave a discernable overdensity of halo samples visible just beyond their boundary. The $E-L_{z}$ plane also shows a distinct change in morphology of the bounding region of the kinematic space, which we mentioned in the previous section is due to there being a fixed velocity budget at a given energy and radius. For the \survey\ sample, where stars at many radii are included, the bounding region is a superposition of many of the aforementioned parabolic curves. These bounding regions tend to get wider as radius increases since the potential scales inversely with distance from the Galactic center. For inner Galaxy samples the bounding parabola is narrow with minima at very negative energies. Conversely, for outer Galaxy stars, the bounding parabola is wide with a minima at higher energy. The superposition of these bounding curves results therefore in a funnel shape which narrows as the energy is decreased. An example of this morphology is shown in Figure~\ref{fig:ELzBoundaries} where the in-plane bounding parabolas for \texttt{MWPotential2014} corresponding to radii 2-15~kpc are shown.

A particularly interesting feature of the \survey\ sample $E-L_{z}$ plane is that there appears to be a region of maximal density which begins at $E \sim -1.3\times10^{5}$ km$^{2}$ s$^{-2}$ and follows an approximate parabolic shape towards larger energies as $\lvert L_{z} \rvert$ increases. These are most obvious in the $\beta=0$ and $0.5$ models, but are also noticable at higher $\beta$. Below this energy there are halo samples, yet they are observed in much lower numbers than at this overdensity. The interpretation of this overdensity is that the majority of the samples in the \survey\ set lie near the Sun, and therefore they tend to be roughly confined by the bounding parabola defined at that location. The occurence of over- and underdensities in $E-L_{z}$ space depending on the underlying radial distribution of samples is something we will explore in the next subsection.

The $\sqrt{J_{R}}-L_{z}$ plane no longer has any indications of the inverse relationship between $J_{R}$ and $L_{z}$ observed in the \solar\ samples and discussed above. Likely the combination of the radial range as well as realistic uncertainties causes all traces of the relationship to disappear. Instead now the halo samples manifest as a broad cloud that is slightly biased either towards $L_{z}$ or $\sqrt{J_{R}}$ depending on the value of $\beta$. The halo samples of the most radially biased models still do exhibit a diffuse cusp along the $L_{z}$ = 0 line. The differences in the $e-L_{z}$ plane tell a similar story. There is no longer as sharp an inverse relationship between eccentricity and $L_{z}$ as was observed in the \solar\ samples, yet the relationship is at least somewhat apparent in contrast with the $\sqrt{J_{R}}-L_{z}$ plane. The higher $\beta$ models still tend to simply cluster more tightly around $L_{z}=0$ and $e=1$, similar to the \solar\ samples. Notably, the disc samples extend all the way to the high-$\beta$ overdensity, and in general completely dominate the prograde half of the plane.

In the action diamond, the disc populations transition from occupying a narrow strip near the prograde corner to dominating the lower right quadrant of the diamond. This large change in area of the disc populations demonstrates a weakness in the action diamond (indeed one shared by eccentricity), which is that ``scaled'' kinematic quantities often confuse orbits that only broadly share certain properties. In this case, when the vertical and radial actions of the hottest thick disc stars become of order comparable with their $L_{z}$ they migrate towards the center and bottom corner of the diamond overlapping with the halo samples, even though the magnitude of their actions can be quite different from those of stars in the halo. The halo samples have the same qualitative trends as are seen in the \solar\ samples. As $\beta$ increases, the samples concentrate near the bottom of the figure. Additionally of note is that very few ergodic samples occupy this region of the diagram, they are more uniformly spread throughout the space and even biased towards the high-$L_{z}$ corners. This means the bottom corner is likely a good region to isolate radially biased halo populations from ergodic populations.

\subsection{The Impact of the Underlying Radial Distribution of Samples on the \texorpdfstring{$E-L_{z}$}{E-Lz} plane}
\label{subsec:RadialBiases}

In the previous section, we noted that the \survey\ samples exhibit an interesting change in morphology in the $E-L_{z}$ plane when compared with the \solar\ samples. In the $E-L_{z}$ plane of the latter (Figure~\ref{fig:SolarKinematics}) the samples were confined by a bounding parabola defined by the escape velocity as a function of changing energy at a fixed Galactocentric radius (this being the location of the Sun in the \solar\ sample). Sample density decreases as energy increases moving away from the bounding parabola. The radially biased samples have a higher energy cusp around $L_{z}=0$, leading to different overall morphology but the same trend still holds. This phenomenon lead us to interpret the apparent overdensity in the $E-L_{z}$ plane of the low-$\beta$ \survey\ samples around $E \sim -1.3\times10^{5}$ km$^{2}$ s$^{-2}$ as being caused by the concentration of the samples near the Sun.

While this observation is not particularly unexpected, it suggests a question: what would happen if the intrinsic radial distribution of samples (i.e. the APOGEE DR16 data which sets the approximate positions of our samples) took a more complicated form? We can investigate this using APOGEE DR16 by augmenting the \survey\ sample with additional samples directed towards the Galactic bulge. Recall that we do not consider any stars for inclusion in the \survey\ sample if they have $-20\degr < \ell < 20\degr$ and $|b| < 20\degr$. Here, just to investigate the radial distribution of samples, we soften that requirement and allow stars lying in this angular range with Galactocentric radius greater than 3~kpc to be included into the \survey\ sample set. Note that this distance from the Galactic center is approximately equal to the distance from the Galactic center enforced by the $20\degr$ angular cut we use to define the \survey\ sample - i.e. $\tan(20\degr) \times 8.178~\mathrm{kpc} \approx 3~\mathrm{kpc}$. This radius is also the radius at which the density of the bulge component of \texttt{MWPotential2014} drops below the density of our stellar halo model when it is normalized to have the mass found by \citet{mackereth20}. This is all to say that we are confident we do not introduce many bulge stars into the sample by including these new data. Separating halo stars from disc stars, sampling kinematics from the DF, and the perturbation by uncertainties are all done in the same manner as described in \S~\ref{sec:DFSamples}. This results in 497 new halo samples.

\begin{figure}
    \centering
    \includegraphics[width=\columnwidth]{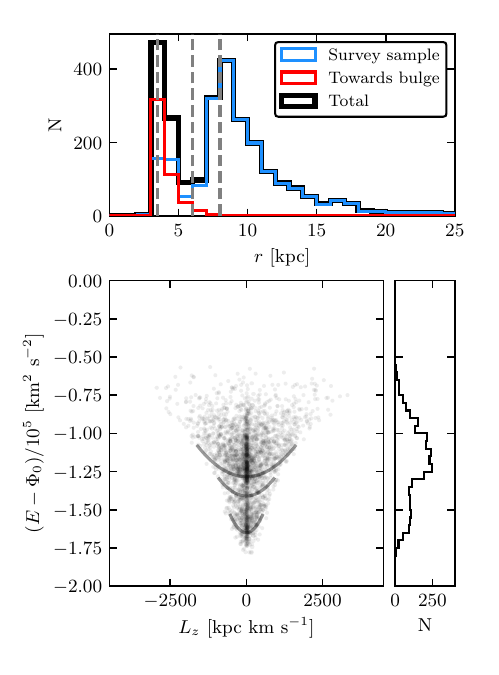}
    \caption{\textit{Top:} Histogram of Galactocentric radii for the \survey\ sample (blue), the additional samples towards the Galactic bulge (red), and their combination (black). \textit{Bottom left:} $E-L_{z}$ plane for \survey\ sample augmented by the additional sources towards the Galactic bulge. \textit{Bottom right:} Histogram of energies for the same sample. Three grey dashed lines in the top panel mark the locations of over and underdensities in the radial distribution at 3.5, 6, and 8 kpc. Three black lines in the bottom panel show sections of the bounding parabolas for those corresponding radii, roughly marking the boundaries between over and underdensities in the $E-L_{z}$ plane.}
    \label{fig:APOGEERadialBias}
\end{figure}

The top panel of Figure~\ref{fig:APOGEERadialBias} shows the radial distribution of the \survey\ samples, the additional samples in the direction of the Galactic center, and their combination. Note that a number of the additional Galactic center samples appear at radii less than 3~kpc. This can be attributed to the perturbation by the APOGEE and \textit{Gaia} uncertainties. The overall distribution of radii is clearly bimodal, with peaks around 3.5 and 8~kpc, and a decrease at 6~kpc. The inner peak is driven mostly by the additional samples from the inner Galaxy pointings, but hints of the overdensity are clearly present in the \survey\ sample. The occurence of this bimodality is undoubtedly caused by the APOGEE selection function, and specifically the location of the survey pointings.

The bottom-left panel of Figure~\ref{fig:APOGEERadialBias} shows the $E-L_{z}$ plane for the fiducial model of the \survey\ sample plus the additional samples towards the Galactic center, 2,695 samples in total (halo samples only, no disc). It is immediately clear that there is an underdensity between $-1.25$ and $-1.5\times10^{5}$~km$^{2}$~s$^{-2}$. This is confirmed by examining the bottom right panel, a histogram of energies, which highlights the overdensity at the aforementioned location. It does not perfectly highlight the overdensity because, as we have established, the density fluctuations occur on roughly parabolic contours, but nontheless its signature is there. Overlaid on the figure are three black lines that are sections of the bounding parabolas corresponding to the three radii marked with grey dashed lines in the top panel: 3.5, 6, and 8~kpc. The bottom and top curves (3.5 and 8~kpc) cradle the two overdensities, while the middle curve at 6~kpc cradles the underdensity. This clearly illustrates that the uneven radial distribution of samples is clearly linked to the appearance of substructure in the $E-L_{z}$ plane.

\section{The Purity of Populations of Interest}
\label{sec:PurityOfPopulations}

One of the most useful questions we might ask is whether we can use our samples to identify which kinematic spaces are best for identifying certain populations that we are interested in separating from a background of contaminants. For the purposes of this work we focus on separating the high-$\beta$ populations from low-$\beta$ and disc populations.

\subsection{Separating High-\texorpdfstring{$\beta$}{beta} Populations from Low-\texorpdfstring{$\beta$}{beta} and Disc Populations}
\label{subsec:SeparatingHighBetaPopulations}

The separation of high-$\beta$ populations from low-$\beta$ populations is currently of special importance, because the two dominant inner-Milky Way halo populations are the \textit{in-situ} low-$\beta$ component and the high-$\beta$ GE component \citep{belokurov18,helmi18,iorio21}. While these two populations do have somewhat unique chemistry \citep{haywood18}, it is not so distinct that that abundances are an efficient means of separating them, such as is the case for separating the disc populations both from one another and from the halo. Building on this sentiment, we do not even know \textit{a priori} the detailed chemical profile of the \textit{in-situ} and GE components so that we might try and separate them using their abundances. Their individual chemistries can only be effectively learned if the populations can be separated by some other means. Additionally, detailed mass modelling in the halo requires some means of tagging different populations of interest. For example \citet{mackereth20} divide the halo into mono-abundance populations. They use a simple eccentricity cut to separate ergodic and high-$\beta$ populations. Here we will delve deeper into the methods for kinematically separating these two major halo populations.

While in the previous section, our focus was on faithfully reproducing the types of samples observed in modern datasets, here we must focus on numerical accuracy. We therefore increase the number of samples in the \survey\ sample set so that the small number of halo stars used to generate the model (2,198) does not impact the fidelity of our results. We do this by generating one hundred different velocity samples at each APOGEE halo star location, rather than just one, for each halo DF with a different $\beta$. This ``augmented'' \survey\ sample set is otherwise processed exactly the same as described in \S~\ref{subsec:TheSurveySamples} and \ref{subsec:SampleKinematics}. For the disc we increase the number of samples by a factor of ten in the same manner. We use these augmented \survey\ samples for the remainder of the paper, and the \solar\ samples remain unchanged.

\begin{figure*}
	\centering
	\includegraphics[width=\textwidth]{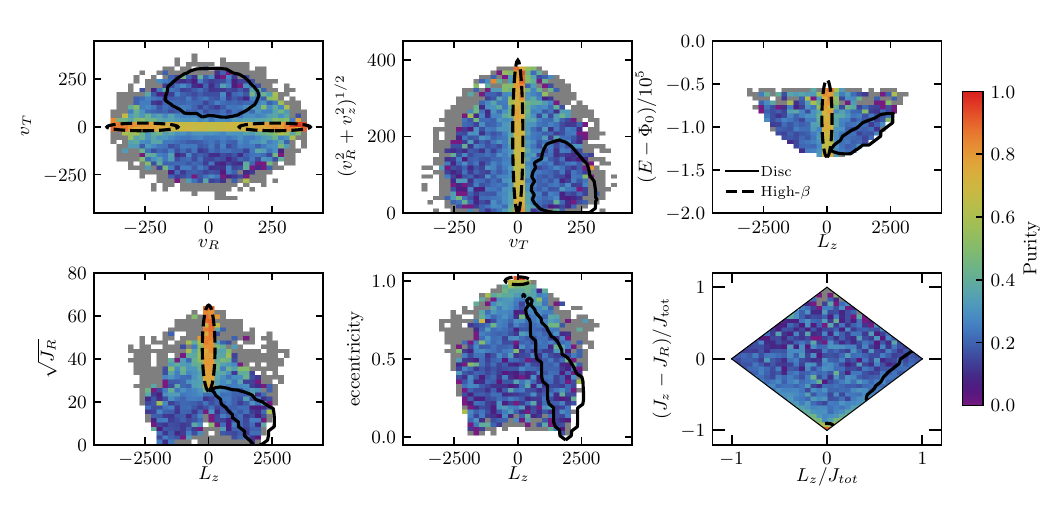}
	\caption{Purity of high-$\beta$ halo samples with respect to low-$\beta$ samples for the \solar\ sample set. The kinematic spaces and units are as described in Figure~\ref{fig:SolarKinematics}. The grey bins are those with fewer than 5 samples in them. In each of the panels the dashed black ellipse(s) show the selection ellipses for the high-$\beta$ sample. The solid black contour is the lowest contour of disc stars from Figure~\ref{fig:SolarKinematics}.}
	\label{fig:SolarHaloPurity}
\end{figure*}

\begin{figure*}
	\centering
	\includegraphics[width=\textwidth]{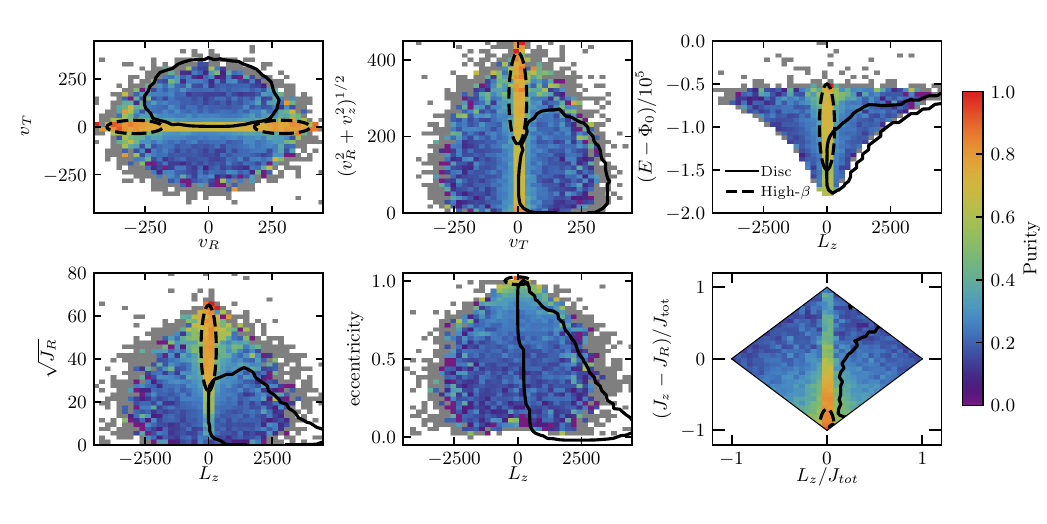}
	\caption{Same as Figure~\ref{fig:SolarHaloPurity} but for the augmented \survey\ sample set. Here, the solid black contour shows ten samples per bin calculated in the same manner as Figure~\ref{fig:SurveyKinematics}.}
	\label{fig:SurveyHaloPurity}
\end{figure*}

We use our fiducial halo model to investigate the purity of ergodic and high-$\beta$ populations; recall that the fiducial halo model consists of a 50:50 mixture of $\beta=0.5$ and 0.9 populations. Figures~\ref{fig:SolarHaloPurity} and \ref{fig:SurveyHaloPurity} show the purity of the high-$\beta$ component of the fiducial model with respect to the low-$\beta$ component for the \solar\ and \survey\ sample sets respectively. Purity is calculated as $N_\mathrm{\beta=0.9}/(N_\mathrm{\beta=0.9}+N_\mathrm{\beta=0.5})$, and the data is binned with 40 bins across each dimension of each kinematic space. Overlaid on each kinematic space is the lowest contour of the disc samples from Figures~\ref{fig:SolarKinematics} and \ref{fig:SurveyKinematics} to indicate where the disc may contribute contaminants to the purity measurement.

Qualitatively, the purity of high-$\beta$ samples follows intuitively from Figures~\ref{fig:SolarKinematics} and \ref{fig:SurveyKinematics}. In the $v_{R}-v_{T}$ and Toomre diagrams the purity is highest along the $v_{T}=0$ axis and at large values of $v_{R}$ and the $v_\mathrm{perp}$ respectively. In the $E-L_{z}$ plane, we see that the purity is highest when $L_{z}$ is zero and at modest to high energies. The $\sqrt{J_{R}}-L_{z}$ plane behaves similarly, with a cloud of high-purity at large values of the radial action near $L_{z}=0$. Eccentricity is peculiar, as we have already noted in that being a ``scaled'' parameter the radially biased populations do not occupy a distinct region of the space. Rather they tend to cluster with greater density near $L_{z}=0$ and $e=1$. This behaviour is necessary to keep in mind when evaluating this space on the merits of high-$\beta$, because one might assume that the best place to isolate radially biased populations is near $e=1$ but at higher $L_{z}$. In reality, the bulk of the sample lies at $L_{z}=0$ but simply exists at lower purity because the ergodic sample also populates that region. The action diamond exhibits the highest purity at the bottom corner where the radial action dominates the action budget. In particular in the \solar\ sample, the region of high-purity is only seen in the few pixels around the apex of the diamond.

\begin{table}
    \centering
    \caption{High-$\beta$ selection ellipse parameters for the \solar\ and \survey\ samples in six kinematic spaces. Ellipses correspond to those in Figures~\ref{fig:SolarHaloPurity} and \ref{fig:SurveyHaloPurity}. See the caption of Figure~\ref{fig:SolarKinematics} for unit conventions. From left to right the columns are: kinematic space, ellipse center point, ellipse semi-major axes, high-$\beta$ completeness, high-$\beta$ purity for halo samples only, and high-$\beta$ purity including disc samples. Note that here, and throughout the paper, we account for the fact that we augmented the \survey\ halo sample by a factor of 100, while we only augmented the disc sample by a factor of 10 when calculating purity including disc samples as contaminants.}
    \begin{tabular}{cccccc}
         Space & Center [x,y] & SMA [x,y] & C & P$_\mathrm{halo}$ & P$_\mathrm{disc}$ \\
        \hline
        \hline
        \multicolumn{5}{c}{\solar\ sample} \\
        \hline
        $v_{R}-v_{T}$        & $[\pm 260,0]$ & [140,20]    & 0.29 & 0.79 & - \\
        Toomre               & $[0,200]$     & [20,200]    & 0.72 & 0.74 & - \\
        $E-L_{z}$            & $[0,-0.9]$    & [200,0.45]  & 0.72 & 0.74 & - \\
        $\sqrt{J_{R}}-L_{z}$ & $[0,45]$      & [250,20]    & 0.69 & 0.78 & - \\
        $e-L_{z}$            & $[0,1]$       & [500,0.025] & 0.62 & 0.83 & - \\
        Action Diamond       & $[0,-1]$      & [0.1,0.1]   & 0.62 & 0.85 & - \\
        \hline
        \multicolumn{5}{c}{\survey\ sample} \\
        \hline
        $v_{R}-v_{T}$        & $[\pm 290,0]$ & [110,35]    & 0.12 & 0.76 & 0.76 \\
        Toomre               & $[0,300]$     & [35,120]    & 0.15 & 0.71 & 0.70 \\
        $E-L_{z}$            & $[0,-1]$      & [300,0.5]   & 0.74 & 0.66 & 0.62 \\
        $\sqrt{J_{R}}-L_{z}$ & $[0,45]$      & [300,20]    & 0.52 & 0.76 & 0.76 \\
        $e-L_{z}$            & $[0,1]$       & [500,0.025] & 0.61 & 0.82 & 0.82 \\
        Action Diamond       & $[0,-1]$      & [0.08,0.3]  & 0.39 & 0.86 & 0.85 \\
        \hline
    \end{tabular}
    \label{tab:HighBetaSelections}
\end{table}

When evaluating how each kinematic space performs in separating populations, it is important to keep in mind the overall density of halo stars. Many regions of each space exhibit spuriously high purity where there are only a few halo samples. We therefore consider only bins with greater than 5 samples in them, and render those bins with fewer than 5 samples grey in Figures~\ref{fig:SolarHaloPurity} and \ref{fig:SurveyHaloPurity}. With this in mind, we have selected elliptical regions in each space where both the purity is large and the density of high-$\beta$ halo samples is reasonable. These are shown as the dashed black lines in Figures~\ref{fig:SolarHaloPurity} and \ref{fig:SurveyHaloPurity}. The parameters of the selection ellipses are given in Table~\ref{tab:HighBetaSelections}. We choose these simple shapes to represent the high-$\beta$ regions because one goal in this analysis is to emphasize applicability to real data. Shapes which are too specific to the populations here may provide errant results when applied to other data where the kinematic properties may be subtly different (e.g., if either the spatial selection or the potential used to derive kinematics changes). Instead, we pick regions which would be mostly insulated from the peculiarities of a specific data set. In each of the six kinematic spaces at least one of the dimensions probes the circularity of the orbit in the disc plane. Following this, each of the selection regions primarily picks out samples with $L_{z}$ or $v_{T}$ approximately equal to zero (the bottom corner of the action diamond necessarily has $L_{z}=0$). The selections then pick out extreme values of the other quantity, which generally trends with the degree of radial bias in a population (specifically $v_{R}$, $v_\mathrm{perp}$, $E$, and $\sqrt{J_{R}}$). The selection regions of the ``scaled'' spaces tend to be concentrated at the extrema of the scaled quantity (eccentricity and $(J_{z}-J_{R})/J_\mathrm{tot}$). The reason for mentioning the overall trends in our choice of selection regions is that these broad properties should suffice to pick out radially biased samples in any data set, not just specifically our DF models. This is important if, for example, we are to be able to confidently apply these selections to real APOGEE DR16 data, something we plan to do in the near future. 

\begin{figure}
    \centering
    \includegraphics[width=\columnwidth]{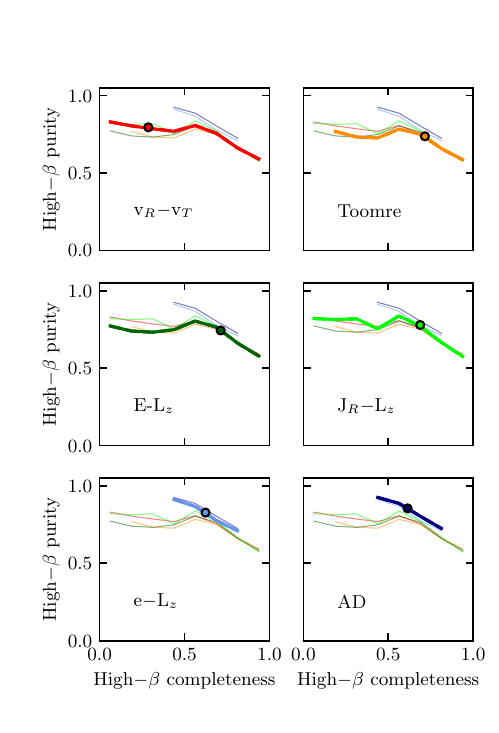}
    \caption{Purity-completeness relation for the fiducial \solar\ sample for each kinematic space (labelled). The bold line in each panel shows the purity calculated only considering halo samples, the solid lines from all other panels are shown faintly in the for context. The filled circle in each panel shows the purity and completeness of the default selections from Figure~\ref{fig:SolarHaloPurity}.}
    \label{fig:SolarCPDiagram}
\end{figure}

\begin{figure}
    \centering
    \includegraphics[width=\columnwidth]{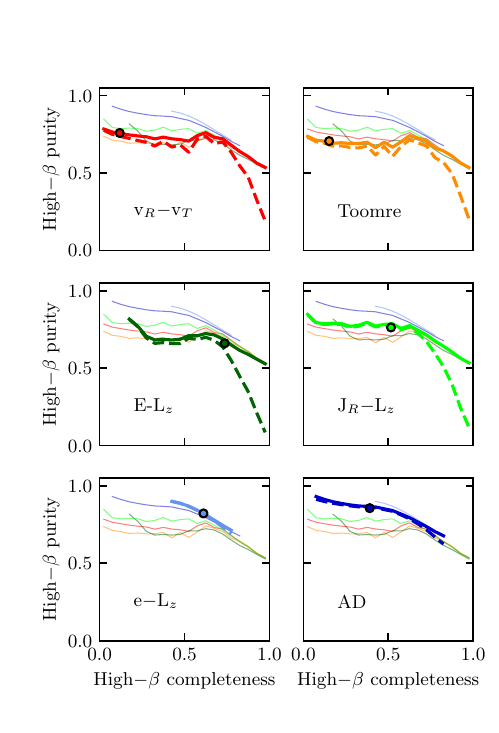}
    \caption{Same as Figure~\ref{fig:SolarCPDiagram} but for the fiducial \survey\ sample. Additionally, the bold dashed line in each panel shows the purity calculated including disc samples. Also see Figure~\ref{fig:SurveyHaloPurity}.}
    \label{fig:SurveyCPDiagram}
\end{figure}

It is still interesting to ask, however, what is the tradeoff between increasing or decreasing the size of the selection ellipses? Intuitively, a larger selection boundary would include more high-$\beta$ samples, but would also include more low-$\beta$ samples which would decrease purity. To explore this, we make purity-completeness relations for each of the kinematic spaces for both the \solar\ and \survey\ sample sets. To do this we consider two sets of scaling factors: $(x_{1},x_{2})$ where each factor takes the form $x_{i} = 2^{y}$. For each of the two scaling factors we consider 25 values of $y$ uniformly spaced on the interval $[-2,2]$, giving a grid of 625 total combinations. For each of these 625 unique pairs of factors we multiply them to the semi-major axes in Table~\ref{tab:HighBetaSelections} to generate a new selection ellipse. We then calculate the high-$\beta$ purity and completeness according to this new selection ellipse. To generate an average purity-completeness relationship we bin the 625 purity measurements in completeness and calculate the mean purity in each bin. The results are shown for the \solar\ and \survey\ sample sets in Figures~\ref{fig:SolarCPDiagram} and \ref{fig:SurveyCPDiagram} respectively. Each of the panels highlights one kinematic space (labelled), the curves for the other kinematic spaces are shown faintly behind to emphasize differences. In the panels of Figure~\ref{fig:SurveyCPDiagram} there is a solid and dashed line that show the purity calculated only considering halo samples (solid) and the purity calculated including both halo and disc samples (dashed). Additionally, in each panel there is a colored circled which shows the purity and completeness of the default boundaries shown in Figures~\ref{fig:SolarHaloPurity} and \ref{fig:SurveyHaloPurity}.

For the \solar\ samples, the purity-completeness relations are similar for quantities which are similar to one another. $v_{R}-v_{T}$ and the Toomre diagram are similar, $E-L_{z}$ and $\sqrt{J_{R}}-L_{z}$ are similar, and $e-L_{z}$ and the action diamond are similar. The action diamond appears to be the superior space, achieving a higher purity at a given completeness than any other kinematic space. At large completeness the other kinematic spaces are more-or-less tied with one another. The apparant bimodalities (i.e. the peaks near completeness of 0.6) in the purity-completeness relations for the absolute kinematic spaces is just an artifact of how we have visualized the results of this analysis. In general, the linear trend at higher completeness in these spaces captures the variation in the rotation-sensitive quantity ($v_{T}$ and $L_{z}$) when the anisotropy-sensitive quantity ($v_{R}$, $v_\mathrm{perp}$, $E$, and $\sqrt{J_{R}}$ is scaled to a large range. At lower completeness the same trend holds when the radial quantity is scaled to small values. These two trends emerge because of the logarithmic manner in which we generate the scaling factors. A general trend for these four spaces is as follows: at a given ellipse size for the anisotropy-sensitive quantity, a narrower selection for the rotation-sensitive quantity produces higher purity and lower completeness. For the action diamond and $e-L_{z}$ the relations are simple because the high-$\beta$ samples occupy a single locus in these spaces. In general the only takeaway from this analysis applied to the \solar\ samples is that the action diamond appears to outperform the other spaces. We are wary of making any other statements because the simplicity of our \solar\ samples clearly incites slightly pathological results in this specific analysis. 

For the \survey\ samples the differences between the kinematic spaces are more interesting. The action diamond and $e-L_{z}$ plane have very similar purity-completeness relations, with the action diamond having lower completeness and higher purity, and the $e-L_{z}$ plane having higher completeness and lower purity. Both of these spaces have little contamination from disc samples (solid and dashed lines are similar). These two scaled spaces appear superior to the other four spaces under consideration. A close third though appears to be the $\sqrt{J_{R}}-L_{z}$ plane, which has slightly lower purity, with the other three spaces at the lowest purities. The shapes of the purity-completeness relations for the velocity planes appears to be slightly peaked at modest completeness. These modest peaks represent selection ellipses that span the whole range of $v_{R}$ and $v_\mathrm{perp}$ respectively. With these selections you would greatly increase the completeness while keeping the purity relatively constant, however the purity including disc stars drops drastically. Overall the $v_{R}-v_{T}$ plane achieves slightly higher purity than the Toomre diagram, likely confirming earlier suspicions that the inclusion of $v_{z}$ in the Toomre diagram's $v_\mathrm{perp}$ actually serves to confuse between low- and high-$\beta$ samples. Finally, the $E-L_{z}$ plane achieves the lowest purities, slightly below the velocity planes, but it does so with quite high completeness. The best kinematic space appears to be the action diamond if purity is of interest. If completeness is of interest, the $e-L_{z}$ plane achieves high completeness while also keeping the purity relatively high.

For both the \solar\ and \survey\ samples we also consider composite selections combining the selection ellipses of multiple kinematic spaces. Intuitively, one might expect that the best composite selection would be to combine a scaled space, such as the action diamond or $e-L_{z}$ with $\sqrt{J_{R}}-L_{z}$ or $v_{R}-v_{T}$. The scaled space selects for stars on very radial orbits, while the other space selects for stars with radial motions of a large magnitude. We find only very minor improvements by combining the action diamond or $e-L_{z}$ selections with the $\sqrt{J_{R}}-L_{z}$ or $v_{R}-v_{T}$ selections for the \survey\ samples. The improvement in purity is at most 0.01, while it tends to reduce completeness by 0.1 to 0.2. For example, when combining the action diamond selection with the $\sqrt{J_{R}}-L_{z}$ selection purity remains the same (compared to just the action diamond), and completeness falls from 0.39 to 0.33. One positive aspect of combining selections is it does tend to erase any disparity between the purity with and without disc samples. The story is the same for the \solar\ samples. With all of this information in mind, the best composite selection (for the \survey\ samples) appears to be the union of the action diamond and $e-L_{z}$ selections. When compared to just the action diamond selection, the purity without (with) disc contaminants is 0.86 (0.86) and the completeness remains constant at 0.39. However this selection is essentially the same as using the action diamond alone.

This analysis does have a few restrictions. First, we impose the use of ellipses because, as we argued above, they are simple selection shapes that can be confidently applied to real data. Second, we fix the location of the ellipses in each kinematic space. We also explored an alternative method for gauging purity-completeness relations for the \solar\ and \survey\ sample sets which does not assume a shape for the selection region, and instead works on a pixel-by-pixel basis. The advantage of this approach is that it examines purity and completeness in an unbiased manner. The weakness is that any selection boundaries based on this individual pixel approach are highly sensitive to the exact kinematic configuration of these data. With this in mind we would not be confident using this method to define selection boundaries for extracting high-$\beta$ stars from real data. The ellipse based method, while imperfect, is sufficiently general that it should work well on real data.

\subsection{Eccentricity as a Single Metric for Separating the High-\texorpdfstring{$\beta$}{beta} and Low-\texorpdfstring{$\beta$}{beta} Halo}
\label{subsec:EccentricityHaloSeparation}

In the analysis presented in the previous section, we found that good purity of high-$\beta$ halo samples with respect to low-$\beta$ halo samples and disc samples can be achieved using eccentricity only if a very narrow region around $e=1$ and $L_{z}=0$ is selected. Outside of this region confusion between high-$\beta$ samples and these contaminants is high. The reasons for needing this very specific selection criterion is as follows: first, that eccentricity is a ``scaled'' quantity, meaning two physically distinct orbits that share a common ratio of apocenter to pericenter have the same eccentricity; second, that as orbits become increasingly radial, eccentricity approaches unity asymptotically. Indeed, examining the $e-L_{z}$ planes in Figures~\ref{fig:SolarKinematics} and \ref{fig:SurveyKinematics} shows that a reasonable fraction of the sample in the ergodic halo models lies at eccentricity near 1.

\begin{figure}
    \centering
    \includegraphics[width=\columnwidth]{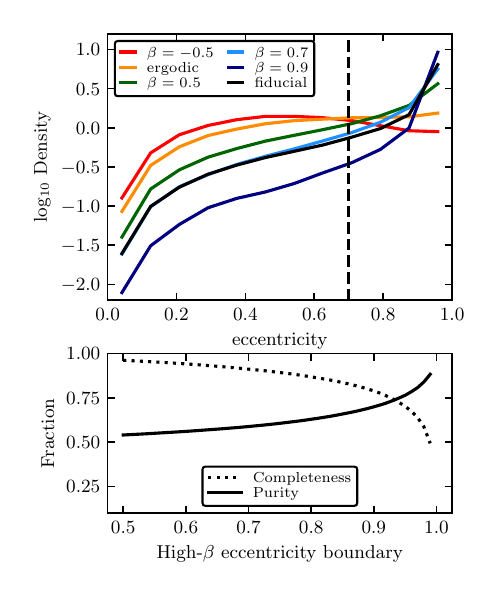}
    \caption{\textit{Top:} eccentricity distribution for the halo samples of each of the six models in the \survey\ sample set. The dashed line marks eccentricity=0.7, a traditional boundary between the ergodic and high-$\beta$ halo populations. \textit{Bottom:} purity and completeness of the high-$\beta$ samples with respect to the ergodic samples in the fiducial halo model of the \survey\ sample set. The quantities are expressed as a function of the cutoff boundary in eccentricity.}
    \label{fig:EccentricityDistributions}
\end{figure}

In the literature, eccentricity is commonly used as a metric to both describe \citep{myeong19,mackereth19a}, and separate \citep{mackereth20,naidu20} low- and high-$\beta$ halo populations. Oftentimes eccentricity is used by itself, as a single parameter without the help of $L_{z}$ or another parameter that selects for tangential orbits. In order to investigate how well these sorts of selection criteria work, we plot the eccentricity distribution function for the halo samples of each of the six models in the \survey\ sample set in the top panel of Figure~\ref{fig:EccentricityDistributions}. We see, as expected, that the peak of the distribution moves towards larger eccentricities as $\beta$ increases from $-0.5$ to $0.9$. Interestingly, we see that the ergodic and $\beta=-0.5$ models have a non-negligible number of samples with very high eccentricity. The ergodic model in particular has a nearly uniform density of samples between eccentricities 0.4 and 1, with a slight peak approaching eccentricity of 1. This suggests that while high eccentricity cuts do a good job of preferentially picking out high-$\beta$ stars, there is a significant amount of contamination by the well-populated high-eccentricity tail of the low-$\beta$ populations.

To quantify this confusion, we plot the completeness and purity of the high-$\beta$ subsets with respect to the low-$\beta$ subset of the fiducial \survey\ sample in the bottom panel of Figure~\ref{fig:EccentricityDistributions}. These quantities are expressed as a function of a cutoff eccentricity, below which we do not consider any samples. The curves are constructed in this way: For each cuttoff eccentricity we take the samples of the fiducial model which have greater eccentricity and calculate purity as $N_\mathrm{\beta=0.9}/(N_\mathrm{\beta=0.9}+N_\mathrm{\beta=0.5})$ and completeness as $N_\mathrm{\beta=0.9}/N_\mathrm{\beta=0.9,tot}$ where $N_\mathrm{\beta=0.9,tot}=$ is 50 per cent of the total number of halo samples in the fiducial model, as per its definition. For smaller eccentricity cuttoffs $\lesssim 0.8$ the purity and completeness vary in an inverse, yet linear fashion. For higher values of the cutoff eccentricity the purity rises sharply while the completeness drops sharply.

Notably, when the eccentricity cuttoff is 0.7 the purity of the high-$\beta$ sample is only 0.59, with the completeness being 0.91. This is a remarkable amount of confusion between these two samples. Recall that the high-$\beta$ purity of the entire fiducial sample is 0.5 by definition, and an eccentricity threshold of 0.7 does little to improve that. When the eccentricity cuttoff is increased to 0.9, the purity rises to 0.7 and the completeness is reduced to 0.79. Even at the largest value we consider for the eccentricity cuttoff, 0.99, the purity only reaches 0.88 (roughly in accordance with our variation of ellipses analysis in the previous section), and the completeness is reduced to 0.49. Note that while this purity and completeness appear superior to those of the action diamond we are wary of advocating for a selection boundary at extreme eccentricity lest it be too specific to our models, but it may be indicative of a good option for certain analyses. Nonetheless these results clearly indicate that care need be taken when separating high-$\beta$ and ergodic populations on the basis of eccentricity alone.

\section{Discussion}
\label{sec:Discussion}

The data and results we present in this paper are clearly useful to discuss in the context of a large fraction of the literature on Milky Way stellar dynamics done in the \textit{Gaia} era, specifically those which focus on the stellar halo. Here we focus on discussing the takeaways we deem are most salient, and how they relate to the wide body of literature.

\subsection{The Validity and Applicability of our Models}

First, we must assess how good the models we have presented are at representing the Milky Way halo as we understand it to be today. From a theoretical perspective, our models are clearly a simplification. For example, it is now well-known that the Milky Way stellar halo is effectively described by a metallicity-dependent triaxial power law density profile \citep[e.g.,][]{mackereth20}. Our models at least have a power law index and cuttoff radius which match observations. But more complicated models are limited by the availability of usable DFs. Additionally, the Milky Way stellar halo is unlikely to currently be in a completely relaxed state, with perturbations from nearby massive dwarf galaxies in various stages of merging \citep{garrow20}, and specifically the Large Magellanic Cloud \citep{erkal19}. Fortunately for us these effects are greatest in the outer halo, with dynamical times in the inner halo where our sample is concentrated being short enough such that these perturbers do not meaningfully impact the equilibria.

A better method of addressing the validity of our models is to take an empirical approach and compare the kinematics of our samples with real observed stellar kinematics. For example, when considering the fiducial \survey\ model, the $v_{R}-v_{T}$ planes of \citet{belokurov18,fattahi19,naidu20}, the Toomre diagram of \citet{helmi18,koppelman19b}, the $E-L_{z}$ plane of \citet{helmi18,koppelman19b,naidu20}, and the action diamond of \citep[normalized to be a circle, ][]{naidu20} all show similarities with the samples produced by our models. Note that there can and should be disimilarities between these kinematics, which arises from the use of different gravitational potentials to calculate the kinematic properties, and varying underlying positional distributions. But overall, on the basis of kinematic properties, our samples are largely consistent with what is seen in real observational data.

Another detail we must address is how well these types of DFs represent the high-$\beta$ stellar halo associated with the debris of GE. This is important given that we spent a great deal of time in \S~\ref{sec:PurityOfPopulations} presenting regions of phase space that uniquely identified high-$\beta$ samples with high purity, with the implication that if we applied these to real data they would serve to identify GE stars. Generalized DFs corresponding to merger debris located in a stellar halo have not been formulated. This is mostly because the orbit and mass ratio of the merging galaxy would play a major role in determining the resulting DF. It is reasonable to assume, however, that the corresponding DF should be somewhere between that of a thin tidal stream occupying a distinct locus in action space \citep[e.g.][]{bovy14}, and the spherical DFs we have used here. It should also be the case that longer timescales and larger mass ratio mergers tend to make the DF of the debris resemble the latter, rather than the former. With this in mind we believe it reasonable to assume that the GE merger debris should resemble a roughly spherical, halo-like DF. This assertion is supported by a wide range of evidence, including that: the GE debris is distributed over the whole sky \citep[see Figure 12 in ][]{helmi20}, the GE debris can be described by a triaxial power law density profile \citep{mackereth20}, and that measured stellar ages indicate GE was accreted roughly 10~Gyr ago \citep{montalban20}, giving the debris ample time to relax.

Additionally, we can look to simulations to further validate our models, and specifically to support our use of high-$\beta$ halo DFs to model the GE remnant. \citet{amorisco17} presents the kinematic properties of debris from a number of simulated mergers parameterized by the mass ratio and orbit circularity of the infalling satellite. They show that the deposited stellar debris has a power law density profile, with some appearance of a cuttoff radius around a few halo scale radii, and the degree of radial bias is high. These facts together support the use of a spherical, radially biased halo DF to model the remnants. \citet{jean-baptiste17} present a series of 1:10 merger simulations which show that the $E-L_{z}$ distribution of the deposited stars is broadly consistent with our high-$\beta$ halo models. They show cusps around $L_{z}=0$ and are roughly bound by extended funnel shaped boundaries. Most of these mergers have a significant degree of angular momentum to them though, which manifests in the skewed appearance of the $E-L_{z}$ distribution.

\subsection{Selection Criteria for \textit{Gaia}-Enceladus}

In \S~\ref{sec:PurityOfPopulations}, we explored the feasibility of using kinematics alone to separate high-$\beta$ populations from low-$\beta$ and disc contaminants within the framework of our fiducial model. We found that most kinematic spaces had regions of high purity, and using well-informed selections, we were able to achieve simultaneously high purity ($\gtrsim 0.7$) and relatively high completeness ($\gtrsim 0.2$). Our main goal in providing these assessments of purity and completeness is to inform and enable identification of GE stars on the basis of kinematics alone. Chemical and stellar properties of GE constituent stars are of significant interest currently, as they tell us a great deal about the progenitor dwarf galaxy. Since it is not obvious \textit{a priori} what these properties are in the context of the broader stellar halo, it is necessary to develop accurate methods to extract GE stars from larger samples of halo and disc stars. It is for this reason that throughout this paper, we have placed more emphasis on achieving high purity when it comes to separating low- and high-$\beta$ populations, than on completeness. In general we recommend using the action diamond, which achieves a high purity. If completeness is of interest we recommend the $e-L_{z}$ plane, which achieves high completeness while keeping purity high.

The selection criteria we present are based upon a number of key assumptions. First, we assume that the ratio of high-$\beta$ to low-$\beta$ stars in the halo is 1:1. While it is clear that these components exist in roughly equal proportion near the solar neighbourhood \citep{belokurov18,lancaster19,iorio21}, the exact ratio is uncertain, and is subject to variation with radius \citep{iorio21}. We explore possible variations in this ratio by repeating our analysis with high- to low-$\beta$ ratios of 3:7 and 7:3. We find that this variation does not significantly change where we would place the selection ellipses for any kinematic space. As one might expect, the purity of the high-$\beta$ selection varies with the fraction of high-$\beta$ stars. For example, the purity without disc contamination of our action diamond selection applied to the \survey\ sample drops to 0.70 when the high- to low-$\beta$ ratio is 3:7, and increases to 0.93 when it is 7:3.

Another assumption built into our model is the use of the \texttt{MWPotential2014} potential of \citet{bovy14}. Notably, this potential has light dark matter halo (virial mass of $8\times10^{11}$~\Msun) when compared to other Milky Way potentials used in the literature. We repeat our analysis using the potential of \citet{mcmillan17}, which has a heavier dark matter halo ($1.2\times10^{12}$~\Msun virial mass) and a slightly larger circular velocity at the location of the sun ($\sim 233$~km~s$^{-1}$). We find next to no differences in our findings using this potential, save for an overall shift in the magnitude of the energies (reflecting the heavier potential) and a barely perceptible increase in the extent of the distribution of the radially sensitive parameters ($v_{R}$, $v_{\mathrm{perp}}$, energy, and $\sqrt{J_{R}}$). This suggests that the selection of high-$\beta$ halo stars is agnostic to any choice of reasonable Milky Way potential.

We also emphasize that these selection are derived using our \survey\ sample set, which is based on the positions of APOGEE DR16 data. While we would not expect the distribution of kinematic properties to change drastically were it based on another survey, there would undoubtedly be noticeable changes. We attempt to gauge how resilient our choice of selections are to a change in the radial distribution of the data by examining the kinematics of samples with $R<7$~kpc,  $7>R>9$~kpc and $R>9$~kpc. In terms of the halo samples, we see clear trends regarding the location of the high-purity regions in each kinematic space. The regions tend to have a wider extent in $L_{z}$ with increasing radius, while the extent of radially sensitive quantities (specifically $E$, $e$, $\sqrt{J_{R}}$, and $(J_{z}-J_{R})/J_{tot}$) increases with decreasing radius. The velocity spaces appear largely unchanged. While it does not appear that the ideal center of the selection ellipses should change with varying radius, their ideal extents should change in a manner reflecting these observed trends. With regards to the disc sources, we observe a slightly different trend. Samples at smaller radii have $L_{z}$ distributions which lie closer to the $L_{z}=0$ / $v_{T}=0$ line where the high-purity halo sample region lies. This would allow the high-$\beta$ selection ellipses to stretch towards lower values of the radially sensitive quantity in each kinematic space. The extent of the radially sensitive quantities for the disc samples also changes, but not in a manner that impacts the regions of high halo purity. We do note that the purity and completeness of many of the selections we have discussed in this work does appear to change by up to $\sim~20$ per cent, but we consider it outside the scope of this work to delve into the radial trends of the completeness and purity for each kinematic space and simply leave this as a cautionary note.

In \S~\ref{subsec:SampleKinematics} we asked whether higher dimensional spaces would be useful for studying the types of kinematic populations we consider in this work. There, we argued that there are really three broad classes of kinematic quantities that we consider here: those sensitive to tangential motions ($v_{T}$ and $L_{z}$), those sensitive to radial motions ($v_{R}$, $v_\mathrm{perp}$, energy, and $\sqrt{J_{R}}$), and those sensitive to the shape of an orbit (eccentricity and $(J_{z}-J_{R})/J_\mathrm{tot})$. Within each of these categories the parameters tend to communicate similar sorts of information, meaning their combination would not be useful and only serve to complicate by increasing dimensionality. For the most part this assertion has been borne out by our results. When we tested combining selections for spaces from within these same categories we find that, in general, purity will remain unchanged, while completeness will decrease (likely owing to the increased dimensionality). For example, when combining the selections for the $v_{R}-v_{T}$ and Toomre spaces the completeness and purity without (with) disc contamination are 0.12 and 0.76 (0.76) respectively. This is essentially identical to the selection which uses just $v_{R}-v_{T}$, suggesting that every point within the smaller (less complete) $v_{R}-v_{T}$ selection also lies within the larger (more complete) Toomre selection; their combination adds no new information. As another example, when $\sqrt{J_{R}}-L_{z}$ selection is combined with the Toomre selection the purity is the same as if one just used the $\sqrt{J_{R}}-L_{z}$ selection yet the completeness falls to 0.13 from 0.52. 

When combining the best quantities from each of the three categories: the action diamond and the $\sqrt{J_{R}}-L_{z}$ plane, the purity is the same as the combination of the action diamond with the $e-L_{z}$ plane (0.86), but the completeness falls to 0.33 (again, likely due to increased dimensionality). As mentioned above, we do find that these composite selections tend to erase any disparity between the purity with and without disc contamination. This improvement is marginal however, the purity including disc contamination of the action diamond, for example, only increases from 0.85 to 0.86 when combining it with $\sqrt{J_{R}}-L_{z}$. We do note that combining the action diamond and $e-L_{z}$ selections also increases the purity including disc contamination to 0.86, but keeps the completeness at 0.39. These results indicate that higher dimensional spaces do not appear to do a better job of separating high- and low-$\beta$ halo populations. They do appear to help in separating halo and disc populations, however, yet the selections we create in the scaled kinematic spaces (action diamond and $e-L_{z}$) are relatively immune from disc contamination to begin with. This result makes sense when considering that disc and halo stars do genuinely have a large difference in the magnitude of their radial motions, whereas it is characteristically the shape of the orbit which distinguishes a star from a low- versus high-$\beta$ population. In other words, there are very few stars in halo populations which have radial motions of a large magnitude, yet their orbit is not radial in shape, and so two dimensional spaces suffice to separate the two populations. 

\begin{figure}
    \centering
    \includegraphics[width=\columnwidth]{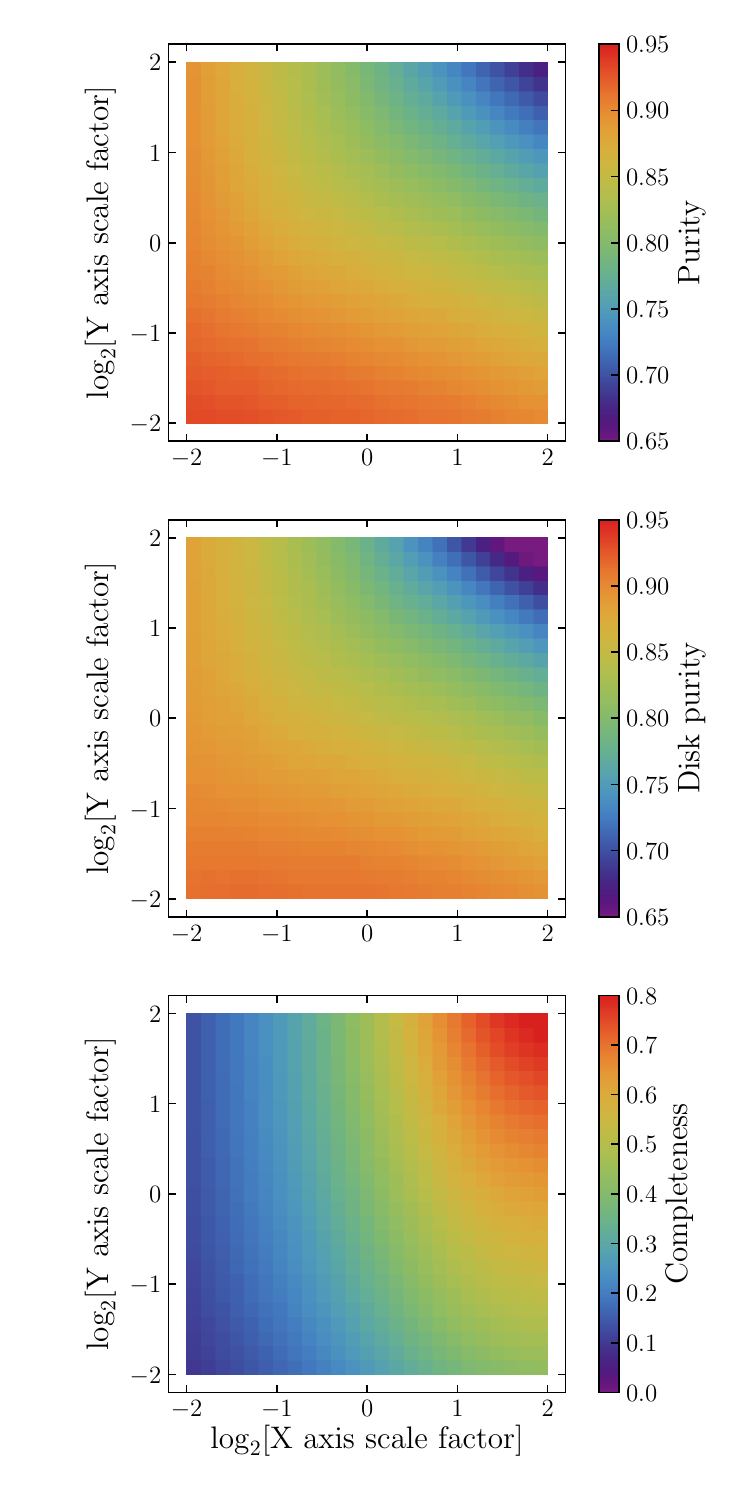}
    \caption{Selection performance as a function of ellipse scale parameters for the \survey\ sample and action diamond space. The three panels show purity (top), purity including disc samples (middle), and completeness (bottom). The scaling factors modify the default action diamond selection ellipse for the \survey\ sample from Table~\ref{tab:HighBetaSelections}.}
    \label{fig:SurveyADCompletenessPurity}
\end{figure}

A final comment, we recognize that our choice of selection regions in each kinematic space is subjective, even though we have endeavoured to provide what we believe are the best choices. We are explicit though in that we favour selections which give a higher purity at a modest expense in completeness, and we also prefer selections which have minimal contamination from disc populations. Nonetheless, other selections which prioritize, for example, completeness over purity may be equally as valid or better for other use cases. In an effort to communicate this tradeoff we plot the purity and completeness as a function of ellipse scale parameters. Figure~\ref{fig:SurveyADCompletenessPurity} shows the performance of the \survey\ sample action diamond selection as the selection ellipse semi-major axes are modified. Purity, purity including disc samples, and completeness are all shown. As the selection ellipse becomes larger the purity decreases while the completeness increases, as expected.

\subsection{Comparison with Literature Selection Criteria}

There already exist selection criteria used in the literature to define the boundaries of GE for chemical and stellar property assays. Here we compare our results with some of the more common selections. For reference, our benchmark selection is the union of our action diamond selection ellipse and $e-L_{z}$ selection ellipse (although as we have mentioned this is very similar to the action diamond selection alone). For the \survey\ sample (the most apt sample set for comparison), the purity of this selection without (with) disc contamination is 0.86 (0.86) and the completeness is 0.39. In order to compare other selections in the fairest manner we evaluate purity and completeness for them using our fiducial \survey\ model. We will quote purity both with and without disc contaminants, but it is important to keep in mind that disc stars may also be efficiently removed from a sample using abundances.

First, we consider the selection criterion of \citet{feuillet20}, which is also used in \citet{matsuno21}, and is defined as $\sqrt{J_{R}} > 30$ kpc$^{1/2}$~km$^{1/2}$~s$^{1/2}$ and $\lvert L_{z} \rvert < 500$ kpc~km~s$^{-1}$. The completeness and purity without (with) disc contaminants of this selection are 0.31 and 0.72 (0.70). Second is the selection of \citet{myeong19}, also used by \citet{monty20} and \citet{cordoni20}, which is defined as $\lvert J_{\phi}/J_\mathrm{tot} \rvert < 0.07$ and $(J_{z}-J_{R})/J_\mathrm{tot} < -0.3$. The completeness and purity without (with) disc contaminants of this selection are 0.58 and 0.82 (0.82). A final selection criterion to consider is that of \citet{naidu20}, based off of \citet{belokurov18} and similar in spirit to \citet{mackereth20}, which is defined as $e>0.7$. We note specifically that \citet{naidu20} apply this criteria to their data only after removing a subset of the halo corresponding to the Sagittarius dwarf and the hot thick disc. The completeness and purity without (with) disc contaminants of this selection are 0.91 and 0.59 (0.37), as per \S~\ref{subsec:EccentricityHaloSeparation}.

Our intent here is not to critique the selection criteria of these studies, but instead to provide an unbiased characterization of their performance in the context of our models. What we see from this comparison, and indeed our study in general, is that when one is kinematically separating high-$\beta$ populations from ergodic populations there is a tradeoff between purity and completeness. If one's interest is in maximal purity of a high-$\beta$ population, then the selection criteria we have provided are best, specifically the union of our action diamond and $e-L_{z}$ selections.

\subsection{Structure Identification}

One of the most remarkable emergent features of our \survey\ models was the appearance of artificial substructure in the distribution of halo samples in the $E-L_{z}$ plane. The substructure is symmetric about the $L_{z}=0$, and appears as a small overdensity of stars at low energies. The fact that this feature appears in our simple halo models defined by a single truncated power law density profile embedded in a sphericalized Milky Way emphasizes that it is entirely spurious. We thoroughly explored this phenomenon in \S~\ref{subsec:RadialBiases}, and found it to be caused by a bimodality in the underlying radial distribution of APOGEE DR16 likely halo stars. Many stars are concentrated near the location of the Sun, about 8~kpc from the center, and many are concentrated also at about 3.5~kpc in radius. The cause of this bimodality is twofold. First, many pointings glance the Galactic bulge (see Figure~\ref{fig:APOGEELocations}), which means there is a concentration of halo stars just on its outskirts which share a galactocentric radius, but have differing galactocentric azimuth. Second, pointings towards the Galactic bulge contribute the most halo stars right on the outskirts of the bulge, as opposed to near the Sun, since observed volume scales with the square of the distance from us, and the density of halo stars should be centrally concentrated.

This is a particularly pertinent result given that the $E-L_{z}$ plane is often used as a means of structure identification \citep[e.g.][]{helmi18,koppelman19b,horta21}, and therefore we strongly recommend that authors consider the selection biases present in their samples when attempting to verify the nature of newly discovered substructures. This is especially important for any survey with a non-trivial spatial selection function, which includes essentially all modern spectroscopic surveys. Furthermore, while the density fluctuations we observe are symmetric about the $L_{z}=0$ axis, we have the clarity of being able to separate halo stars from disc stars. This is not a luxury afforded to most samples, which will also contain contamination from disc stars. These contaminants would potentially hide the symmetric nature of structure that arises thanks to the types of radial selection biases we explore here. This could result in, for example, the density fluctuations being observed in the lower density retrograde half of the plane but not the prograde half where they could be masked by disc contaminants.

We specifically highlight the case of the newly discovered Heracles structure in the inner Galactic halo by \citet{horta21} for two reasons. First, they also use the APOGEE DR16 data that underpins our models. Second, the structure they identify shares many features in common with the artifacts we find. This is to say that Heracles is roughly symmetric about the $L_{z}=0$ axis, and it appears as a separate overdensity of stars at lower energies than the primary group of halo stars. We suggest that the structure be further investigated to confirm that it does not appear as a result of the APOGEE selection function.

In a similar vein, a potentially useful application of the sorts of models we have presented here, is to use them for the differential identification of substructures in both the disc and stellar halo. Because our models use simple, smooth underlying potentials, they can be used to generate phase space realizations of those smooth potentials which can be compared with observations. The advantage of this approach is that it would be relatively simple to account for the types of selection effects we just highlighted, because phase space need only be sampled at the locations where the real data are found. A major challenge of this approach is determining the impact of assuming the nature of the underlying potential, but this could be circumvented to a degree by comparing many different potentials.

\section{Conclusion}

In this work, we have demonstrated a framework for assessing the kinematic properties of the types of composite stellar populations which are observed by modern spectroscopic surveys complemented by \textit{Gaia} kinematics. Our methodology relies on a simple, DF-based approach to generate mock kinematics for both solar-vicinity and APOGEE DR16 data using \texttt{galpy}, and would be simple to replicate for other surveys. We therefore suggest that these sorts of models be created for many different surveys in order that their unique kinematic distributions and underlying selection effects can be properly gauged.

We provide a general overview of many kinematic spaces commonly used in the study of the Milky Way, and specifically the stellar halo. In each kinematic space, we show two sets of models: the \solar\ models which are simple and mimic stellar samples concentrated near the Sun, and the \survey\ models which are more complicated and designed to mimic samples observed with spectroscopic surveys. Each model uses a stellar halo parameterized by the anisotropy: $\beta$, which is a useful quantity for describing the two dominant stellar halo populations in the Milky Way: the low-$\beta$ metal poor halo and the high-$\beta$ GE remnant.

In each kinematic space we use our models to show where disc populations, low- and high-$\beta$ halo populations all lie. We use purity as a metric to specifically assess each space on how well it separates low-$\beta$ halo populations from high-$\beta$ populations, finding that the scaled action diamond, the $\sqrt{J_{R}}-L_{z}$ plane, and $e-L_{z}$ plane achieve highest performance, with a focus on purity. These results assume a 1:1 ratio of low- and high-$\beta$ populations in the stellar halo, and also use APOGEE DR16 data as a set of positions for the phase space samples. Despite these constraints we believe our choice of selection regions in each kinematic space is resilient to changes in these and other underlying assumptions. While the location of the best selection regions is resilient to changes in assumptions, the specific purity and completeness values are liable to change. We therefore recommend that other authors generate their own completeness and purity corrections for their data if the specific values are needed. We briefly explored the effects of survey selection functions on kinematic properties of observed samples, finding that they can result in the appearance of specious structures.

We summarize this work with a number of recommendations for authors to consider as they explore current and future datasets:

\begin{itemize}
    \item The scaled action diamond, $e-L_{z}$ plane, and $\sqrt{J_{R}}-L_{z}$ plane are best for separating high- and low-$\beta$ populations. We provide a number of simple-to-implement selection criteria (see Table~\ref{tab:HighBetaSelections}) for separating high- and low-$\beta$ populations in each kinematic space we study.
    
	\item If high purity is most important when separating low- and high-$\beta$ populations, then we recommend using our action diamond selection. It achieves a purity without (with) disc contamination of 0.86 (0.85) and completeness of 0.39. Combining the action diamond selection with the $e-L_{z}$ selection increases the purity with disc contamination to 0.86, while keeping the completeness constant at 0.39.
    
    \item If high completeness is most important when separating low- and high-$\beta$ populations, then we recommend using our $e-L_{z}$ selection. It achieves a purity without (with) disc contamination of 0.82 (0.82) and completeness of 0.61. We also show the purity and completeness of a wide range of selection ellipse sizes for the action diamond, to illustrate how completeness may be increased for that selection given a tradeoff in purity.
   
    \item If invariance to choice of Milky Way potential is most important when separating low- and high-$\beta$ populations, then we recommend using the $v_{R}-v_{T}$ plane, which outperforms the Toomre diagram.
    
    \item In general, the $E-L_{z}$ plane and Toomre diagram do not do as good a job of separating low- and high-$\beta$ populations. Additionally, we recommend that caution be exercised using eccentricity by itself for this purpose. Commonly used eccentricity cuts for the radially anisotropic halo, such as $e\sim0.7$ only result in a sample with purity of 0.59. Only extreme eccentricity cuts ($e \gtrsim 0.95$) achieve good purity.
    
    \item Be aware of all selection biases in the underlying sample, which can cause spurious structures to appear in kinematic spaces with explicit dependencies on radius, specifically in the $E-L_{z}$ plane. We show that in APOGEE DR16, a bias in the underlying Galactocentric radial distribution of sources results in the appearance of artifacts in the $E-L_{z}$ plane which could be easily mistaken for real substructure.
\end{itemize}

\section*{Acknowledgements}

JMML and JB acknowledge financial support from NSERC (funding reference number RGPIN-2020-04712) and an Ontario Early Researcher Award (ER16-12-061). This work has made use of data from the European Space Agency (ESA) mission
{\it Gaia} (\url{https://www.cosmos.esa.int/gaia}), processed by the {\it Gaia}
Data Processing and Analysis Consortium (DPAC,
\url{https://www.cosmos.esa.int/web/gaia/dpac/consortium}). Funding for the DPAC
has been provided by national institutions, in particular the institutions
participating in the {\it Gaia} Multilateral Agreement. Funding for the Sloan Digital Sky Survey IV has been provided by the Alfred P. Sloan Foundation, the U.S. Department of Energy Office of Science, and the Participating Institutions. SDSS-IV acknowledges support and resources from the Center for High-Performance Computing at the University of Utah. The SDSS web site is \url{www.sdss.org}.

\section*{Data Availability}
 
The APOGEE DR16 data used in this article are available at: \url{https://www.sdss.org/dr16}. The \textit{Gaia} data used in this article are available at: \url{https://gea.esac.esa.int/archive/}.


\bibliographystyle{mnras}
\bibliography{manuscript}



\appendix

\section{Ergodic and anisotropic distribution functions for dark-matter halos and their stellar halos in \lowercase{\texttt{galpy}}}\label{dfappendix}

To model the kinematics of the stellar halo and spherical galactic components more generally, we have implemented various spherical distribution functions in the \texttt{galpy} code. In this Appendix, we provide a brief overview of the implemented functionality and its mathematical basis that serves as a reference for the code. We consider distribution functions $f(E,L)$ embedded in a spherical potential $\Phi(r)$ that are functions of the specific energy $E$ and the total specific angular momentum $L$ and that have either constant orbital anisotropy $\beta$ or anisotropy of the Osipkov-Merritt type. Such models are standard textbook material (e.g., \citealt{binney08}), but for completeness we review the main theory behind them here.

\subsection{Spherical coordinates}

We work in spherical coordinates $(r,\phi,\theta)$, with $\theta$ the polar angle measured from the pole and $\phi$ the azimuthal angle. In terms of the usual cartesian coordinates $(x,y,z)$, the spherical coordinates are given by
\begin{align}
    \nonumber r       &= \sqrt{x^2 + y^2 + z^2}\,,\\
    \phi    &= \mathrm{atan2}(y,x)\,,\\
    \nonumber \theta  &= \pi/2 - \mathrm{atan}(z/\sqrt{x^2+y^2})\,,
\end{align}
and the inverse transformation is
\begin{align}
    \nonumber x  &= r\,\sin \theta\, \cos \phi\,,\\
    y  &= r\,\sin \theta\, \sin \phi\,,\label{eq-sphere-coords}\\
    \nonumber z  &= r\,\cos \theta\,.
\end{align}
In the spherical coordinate frame, the velocities  $(v_r,v_\phi,v_\theta)$ are given in terms of the cartesian velocities $(v_x,v_y,v_z)$ as
\begin{align}
    \nonumber v_r & = \phantom{-}v_x\,\sin\theta\,\cos\phi+v_y\,\sin\theta\,\sin\phi+v_z\,\cos\theta\\
    v_\phi & = -v_x\,\sin\phi\phantom{\,\cos\phi}+v_y\,\cos\phi\\
    \nonumber v_\theta & = \phantom{-}v_x\,\sin\theta\,\cos\phi+v_y\,\sin\theta\,\sin\phi-v_z\,\cos\theta\,,
\end{align}
with inverse transformation
\begin{align}
    \nonumber v_x & = v_r\,\sin\theta\,\cos\phi-v_\phi\,\sin \phi+v_\theta\,\cos\theta\,\cos\phi\\
    v_y & = v_r\,\sin\theta\,\sin\phi+v_\phi\,\cos\phi+v_\theta\,\cos\theta\,\sin \phi\label{eq-vel-spher}\\
    \nonumber v_z & = v_r\,\cos\theta\phantom{\,\sin\phi+v_\phi\,\cos\phi}-v_\theta\,\sin\theta\,.
\end{align}
When working with the velocities, we will also express them using a spherical coordinate system $(v,\psi,\eta)$ of their own, writing the velocities as
\begin{align}
    \nonumber v_{r} & = v \cos\eta\,, \\
     v_{\theta} & = v \sin \eta \cos \psi \label{eq-sphervels}\\
    \nonumber v_{\phi} & = v \sin \eta \sin \psi\,.
\end{align}
In these coordinates, the specific energy is
\begin{equation}\label{eq:E-as-v}
    E = \Phi(r) + {v^2 \over 2}\,,
\end{equation}
while the total specific angular momentum is given by
\begin{equation}
    L = rv\sin \eta\,,
\end{equation}
A distribution function $f(E,L)$ is therefore only a function of $(r,v,\eta)$ and does not depend on the two spatial angles $(\phi,\theta)$ or the velocity angle $\psi$.

The orbital anisotropy $\beta$ for our DFs is defined as
\begin{equation}
    \beta = 1-{\sigma_\theta^2+\sigma_\phi^2 \over 2\sigma_r^2} = 1-{\sigma_t^2 \over 2\sigma_r^2}\,,
\end{equation}
where $\sigma_{r,\phi,\theta}$ are the velocity dispersions in the radial, azimuthal, and polar directions and all quantities are measured at a given radius $r$; the second equality uses the velocity dispersion in the tangential velocity $v_t = v\,\sin\eta$.

\subsection{Distribution functions}

As is usual, we work in terms of the relative potential $\Psi = -\Phi+\Phi(\infty)$ and the specific binding energy $\mathcal{E} =  -E +\Phi(\infty) = \Psi -\frac{1}{2}\,v^2$. Bound orbits then have $0 < \mathcal{E} < \mathcal{E}_{\mathrm{max}}$, where $\mathcal{E}_{\mathrm{max}}$ is the maximum binding energy given by $\mathcal{E}_{\mathrm{max}} = \Psi(0) = -\Phi(0)+\Phi(\infty)$.

The first type of spherical DF that we consider are the \emph{ergodic DFs}
\begin{equation}
    f(\mathcal{E},L) \equiv f(\mathcal{E})\,.
\end{equation}
Because $\mathcal{E}$ does not depend on $\sin\eta$ (Equation \ref{eq:E-as-v}), such DFs have $\beta=0$ and are therefore isotropic. For a given density $\rho(r)$ embedded in a spherical potential $\Psi$, the DF is given by the Eddington inversion \citep{Eddington16a}
\begin{equation}\label{eq-eddington-alt}
    f(\mathcal{E}) = \frac{1}{\sqrt{8}\,\pi^2}\,\frac{\mathrm{d}}{\mathrm{d}\mathcal{E}}\,\int_0^\mathcal{E}\mathrm{d}\Psi\,\frac{1}{\sqrt{\mathcal{E}-\Psi}}\,\frac{\mathrm{d}\rho}{\mathrm{d}\Psi}\,,
\end{equation}
or, equivalently,
\begin{equation}\label{eq-eddington}
    f(\mathcal{E}) = \frac{1}{\sqrt{8}\,\pi^2}\,\left[\int_0^\mathcal{E}\mathrm{d}\Psi\,\frac{1}{\sqrt{\mathcal{E}-\Psi}}\,\frac{\mathrm{d}^2\rho}{\mathrm{d}\Psi^2} +\frac{1}{\sqrt{\mathcal{E}}}\,\frac{\mathrm{d}\rho}{\mathrm{d}\Psi}\Bigg|_{\Psi=0}\right]\,.
\end{equation}
Note that the density $\rho(r)$ does not have to be related to the potential $\Psi$ through the Poisson equation; this is the case for the anisotropic DFs below as well. We will only deal with DFs where the last term is zero.

DFs with a constant, but non-zero, value of the anisotropy $\beta$ can be obtained using the form
\begin{equation}\label{eq-constantbetadf-form}
    f(\mathcal{E},L) \equiv L^{-2\beta}\,f_1(\mathcal{E})\,.
\end{equation}
As shown by \citet{Cuddeford91a} (see also \citealt{An06a} for a clearer statement of this result), for a given density $\rho(r)$ embedded in a spherical potential $\Psi$, the DF is given by the inversion 
\begin{equation}\label{eq-anidf-constbeta-generalsol}
\begin{split}
    f(\mathcal{E},L) =  & {2^\beta\,L^{-2\beta} \over (2\pi)^{3/2}\,\Gamma(1-\alpha)\,\Gamma(1-\beta)}\\&\ \times \left[\int_0^\mathcal{E}\mathrm{d}\Psi\,{1\over (\mathcal{E}-\Psi)^\alpha}\,{\mathrm{d}^{m+1} \over \mathrm{d} \Psi^{m+1}}\Big(\rho(\Psi)\,[r(\Psi)]^{2\beta}\Big)\right.\\& \quad \left. +{1\over \mathcal{E}^\alpha}\,{\mathrm{d}^{m} \over \mathrm{d} \Psi^{m}}\Big(\rho(\Psi)\,[r(\Psi)]^{2\beta}\Big)\Big|_{\Psi=0}\right]\,,
\end{split}
\end{equation}
where $m = \lfloor 3/2-\beta \rfloor$ and $\alpha = 3/2-\beta-m$. For half-integer values of $\beta$, the DF is simply given by
\begin{equation}\label{eq-anidf-constbeta-halfinteger}
    f(\mathcal{E},L) = {L^{-2\beta} \over 2\pi^2\,(-2\beta)!!}\,{\mathrm{d}^{3/2-\beta} \over \mathrm{d} \Psi^{3/2-\beta}}\Big(\rho(\Psi)\,[r(\Psi)]^{2\beta}\Big)\,,\quad 1/2-\beta \in \mathbb{N}\,.
\end{equation}
where $(-2\beta)!! = 2^{1/2-\beta}\,\Gamma(1-\beta)/\sqrt{\pi}$.

\begin{figure}
	\centering
	\includegraphics[width=\columnwidth]{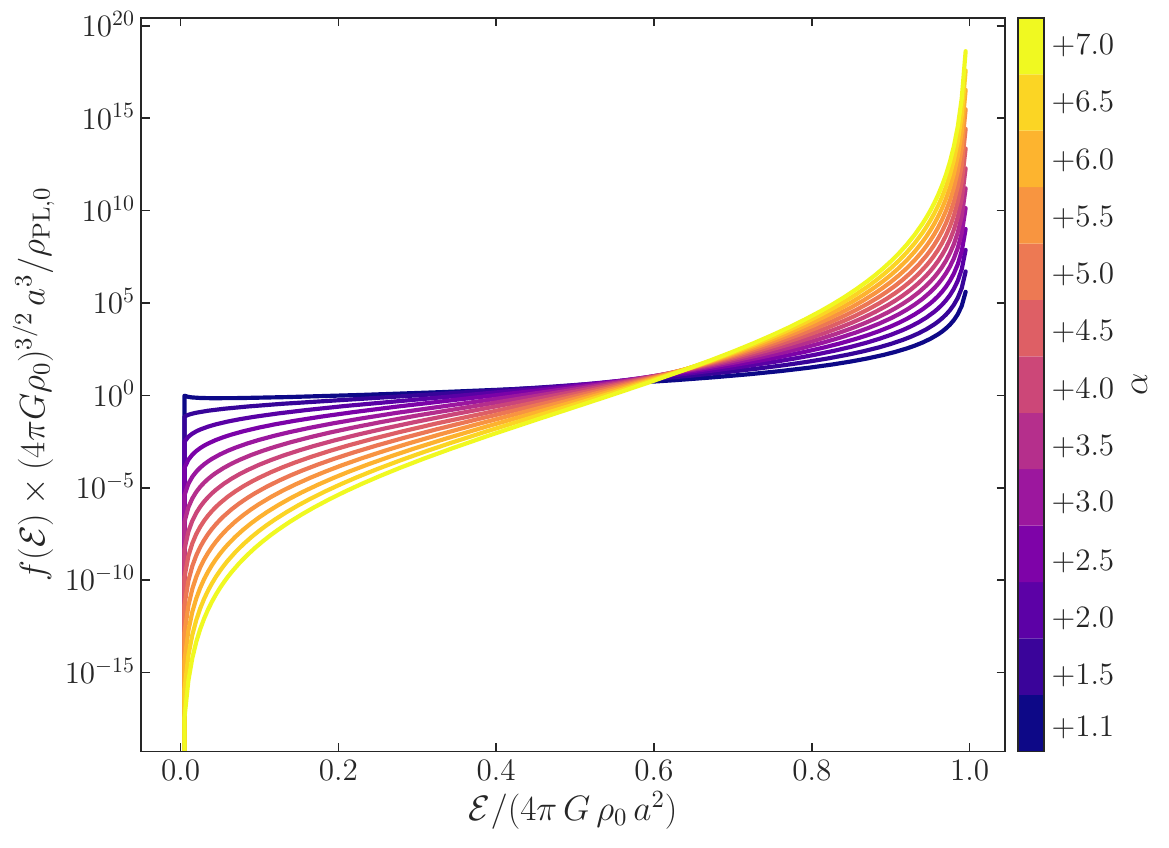}
	\caption{Ergodic distribution function $f(\mathcal{E})$ for a power-law density profile $\rho(r) = \rho_{\mathrm{PL},0} (r/a)^{-\alpha}$ embedded in an NFW potential $\Phi(r) = -4\pi\,G\,\rho_0\,a^2\,\ln(1+r/a)/(r/a)$ in terms of the dimensionless binding energy $\mathcal{E}/(4\pi\,G\,\rho_0\,a^2)$.}
	\label{fig:pl_in_nfw}
\end{figure}

As first proposed by \citet{osipkov79} and \citet{merritt85}, DFs for a given density $\rho$ embedded in a gravitational potential $\Psi$ with radially-varying anisotropy of the form
\begin{equation}\label{eq-anidf-osipkov}
    \beta(r) = {1 \over 1+r_a^2/r^2}\,,
\end{equation}
can be obtained from DFs of the form
\begin{equation}
    f(\mathcal{E},L) \equiv f(Q)\,,
\end{equation}
where
\begin{equation}
    Q = \mathcal{E}-{L^2 \over 2r_a^2}\,,
\end{equation}
with $r_a$ the anisotropy radius (the radius where $\beta = 1/2$). For a given density $\rho$ embedded in a potential $\Psi$, this DF is given by an Eddington-like inversion
\begin{equation}
\begin{split}
    f(Q) =  \frac{1}{\sqrt{8}\,\pi^2}\,& \left[ \int_0^Q\mathrm{d}\Psi\,\frac{1}{\sqrt{Q-\Psi}}\,\frac{\mathrm{d}^2}{\mathrm{d}\Psi^2}\Big[\rho(\Psi)\,(1+r^2[\Psi]/r_a^2)\Big]\right.\\&  \ \left. +\frac{1}{\sqrt{Q}}\,\frac{\mathrm{d}}{\mathrm{d}\Psi}\Big[\rho(\Psi)\,(1+r^2[\Psi]/r_a^2)\Big]\Big|_{\Psi=0}\right]\,.
\end{split}
\end{equation}
This can also be written as
\begin{equation}\label{eq-spherdf-omdf}
\begin{split}
    f(Q) = f^{\beta=0}(Q)+\frac{1}{\sqrt{8}\,\pi^2\,r_a^2}\,&\left[\int_0^Q\mathrm{d}\Psi\,\frac{1}{\sqrt{Q-\Psi}}\,\frac{\mathrm{d}^2}{\mathrm{d}\Psi^2}\Big[\rho(\Psi)\,r^2(\Psi)\Big]\right.\\ & \ \left.+\frac{1}{\sqrt{Q}}\,\frac{\mathrm{d}}{\mathrm{d}\Psi}\Big[\rho(\Psi)\,r^2(\Psi)\Big]\Big|_{\Psi=0}\right]\,,
\end{split}
\end{equation}
where $f^{\beta=0}(Q)$ is the ergodic DF from Equation \eqref{eq-eddington} evaluated at $\mathcal{E} = Q$. Because the square brackets in the second term do not depend on $r_a$, Osipkov-Merritt DFs for a given $(\rho,\Psi)$ pair for all values of $r_a$ can be efficiently computed from two one-dimensional quadratures that do not depend on $r_a$.

\subsection{Numerical implementation}

\begin{figure}
	\centering
	\includegraphics[width=\columnwidth]{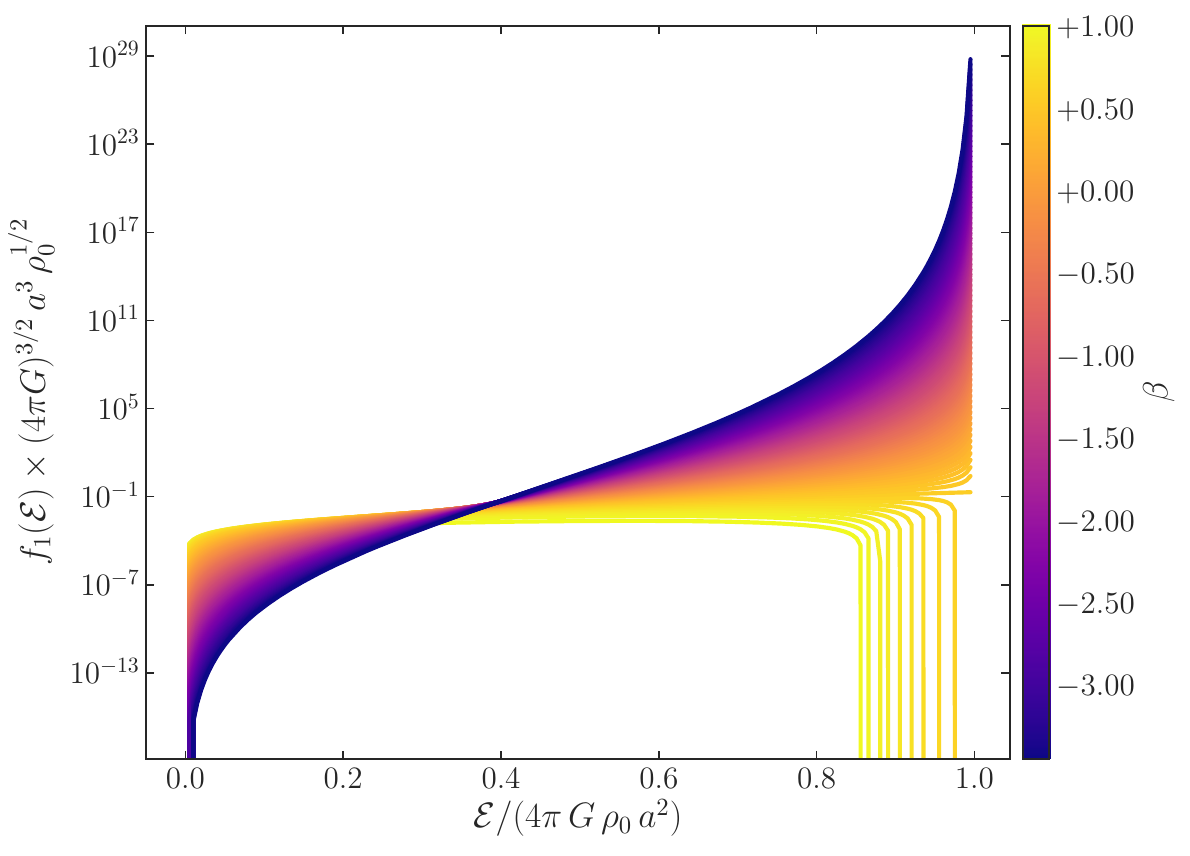}
	\caption{Energy part $f_1(\mathcal{E})$ of the self-consistent distribution function for an NFW profile with constant anisotropy $\beta$ for a dense grid in $\beta$. At $\beta > 0.5$, the distribution function becomes increasingly unphysical.}
	\label{fig:constantbeta_nfw}
\end{figure}

All of the DFs given in the previous subsection can be computed through one-dimensional quadratures over the relative potential $\Psi$. The only exception to this are the half-integer-$\beta$ constant-anisotropy DFs from Equation \eqref{eq-anidf-constbeta-halfinteger}, which do not require a numerical quadrature and are therefore straightforward to compute. The numerical integrals are difficult to compute for two reasons: (i) the integral is over $\Psi$, but the density is expressed as a function of $r$ and the integrand is therefore most easily evaluated as a function of $r$, and (ii) the integrand diverges integrably at the upper limit. To deal with difficulty (i), we rewrite all integrals to be over $r$ rather than $\Psi$; this requires the Jacobian $\mathrm{d} \Psi / \mathrm{d} r$, which is simply equal to the gravitational field. We also need to transform the limits of the integration: the lower limit of $\Psi = 0$ is the maximum radius $r_{\mathrm{max}}$ of the system and is typically infinity, the upper limit $\Psi = \mathcal{E}$ is transformed to $r$ by interpolating the function $r(\Psi)$. Because the integral after this transformation is well-behaved, the numerical error is easily dominated by how well this interpolation works and we therefore use a fine, logarithmically-spaced grid of $r$ values to determine $r(\Psi)$. After the transformation, the integral extends from $r(\mathcal{E})$ to $r=r_{\mathrm{max}}$ and when $r_{\mathrm{max}}=\infty$, this means that special handling of the integral is required at both limits: dealing with the integrable divergence at one end and an infinite limit at the other end. To deal with this, we split the integral at twice the lower limit and then use standard transformation methods to deal with the divergence and the infinite limit (e.g., \citealt{Press07a}). 

The implementation of the Eddington inversion in Equation \eqref{eq-eddington-alt} requires the second derivative of the density with respect to $\Psi$, while the constant-anisotropy DFs require derivatives of $\rho\,r^{2\beta}$ and the Osipkov-Merritt DFs need derivatives of $\rho\,r^{2}$. To compute these, we first of all rewrite them all as derivatives with respect to $r$ using the chain rule; this introduces derivatives of the relative potential that can be expressed in terms of the gravitational field and higher order derivatives of $\Psi$. For ergodic and Osipkov-Merritt DFs we only need at most two derivatives of either $\rho$ or $\Psi$ and to compute these, we explicitly add the first and second derivative of the density to all \texttt{galpy} spherical potentials; second derivatives of the potential already existed. Depending on how low $\beta$ goes, we need arbitrary derivatives of $\rho$ and $\Psi$ with respect to $r$. Rather than computing these by hand, we use \texttt{jax} \citep{jax2018github}, a Python library that implements automatic differentiation of arbitrary functions built using \texttt{numpy} functions and more and that as such can compute arbitrary derivatives of functions to machine precision. To use \texttt{jax} in our case, we implement $\mathrm{d} (\rho\,r^{2\beta})/\mathrm{d} r$ as well as the radial field $\mathrm{d} \Psi / \mathrm{d}r$ using \texttt{jax}'s version of \texttt{numpy}. Note that for many potentials, the density derivative or the radial field does not involve special functions and it is then not actually necessary to load \texttt{jax}'s \texttt{numpy} implementation. The \texttt{jax} derivatives are also used to compute the necessary derivatives in the half-integer case of Equation \eqref{eq-anidf-constbeta-halfinteger}.

As an example of our numerical implementation, we display the ergodic DF $f(\mathcal{E})$ calculated using Equation \eqref{eq-eddington-alt} for a power-law density profile $\rho(r) = \rho_{\mathrm{PL},0} (r/a)^{-\alpha}$ that is embedded in an NFW potential with scale parameter $\alpha$
\begin{equation}\label{eq-nfw-pot}
\Phi(r) = -4\pi\,G\,\rho_0\,a^2\,\ln(1+r/a)/(r/a)\,
\end{equation}
in Figure \ref{fig:pl_in_nfw} for different values of the power-law exponent $\alpha$. As expected, more centrally-concentrated profiles with higher values of $\alpha$ have more stars on tightly-bound orbits. All of these DFs are physical, that is, do not contain negative regions.

A further example of our numerical implementation is given in Figure \ref{fig:constantbeta_nfw}, which shows the $f_1(\mathcal{E})$ part of the self-consistent DF for an NFW profile with constant anisotropy $\beta$ for a dense grid in $\beta$. The smoothness of the variation of the DF's shape with $\beta$ provides a good indication of the high numerical fidelity of our implementation. At $\beta > 0.5$, an increasingly large region at high binding energy is negative, indicating that there is no physical DF with such high constant anisotropy of the form of Equation \eqref{eq-constantbetadf-form}.

\subsection{Sampling the distribution functions}

To be able to sample six-dimensional phase-space coordinates for a wide range of spherical DFs, we have implemented exact random sampling for all of the DFs described in this Appendix. For all DFs, we sample positions $(r,\phi,\theta)$ by sampling $r$ from the cumulative mass profile, using inverse transform sampling, and sampling random angles on the sphere $(\theta,\phi)$ keeping in mind that while the distribution of $\phi$ is uniform on $ \phi \in [0,2\pi]$, the distribution of $\theta$ is $p(\theta)\propto \sin \theta$ on $\theta \in [0,\pi]$. Here and below, when the inverse cumulative distribution used in inverse transform sampling cannot be computed analytically, we calculate it using interpolating the inverse transform computed on a grid.

Sampling velocities necessarily depends on the anisotropy profile of the DF. For all DFs, we sample velocities in the spherical velocity coordinate system $(v,\psi,\eta)$ from Equation \eqref{eq-sphervels}. For DFs that only depend on $E$ and $L$, the distribution of the angle $\psi$ is uniform on $\psi \in [0,2\pi]$. For ergodic DFs, the distribution of $\eta$ is $p(\eta) \propto \sin \eta$, which can be sampled exactly using inverse-transform sampling using the fact that the cumulative DF can be inverted analytically. For DFs with constant $\beta$, we have that $p(\eta) \propto \sin^{1-2\beta}\eta$, which can also be sampled using inverse transform sampling (although while the cumulative DF can be expressed using special functions, it cannot be inverted analytically). We then sample the magnitude of the velocity from $p(v|r) \propto v^2 f(\mathcal{E})$ or $p(v|r) \propto v^2 f_1(\mathcal{E})$ for non-zero, constant $\beta$ using inverse transform sampling between zero and $v = v_{\mathrm{esc}}(r)$.

Exactly sampling Osipkov-Merritt-type DFs is slightly more difficult and has to our knowledge not been discussed before. We first sample the velocity angle $\eta$ from $p(\eta|r) \propto \sin\eta\,[1+(r/r_a)^2 \sin^2\eta]^{-3/2}$ using inverse transform sampling. We then proceed to sample $v'=v\sqrt{1+(r/r_a)^2\sin^2 \eta}$ from $p(v'|r)\propto v'^2 f(Q)$. Remarkably, the cumulative distribution of $\cos \eta$ can be analytically inverted, so sampling it using inverse transform sampling is straightforward.

In practice, we interpolate the inverse cumulative distributions for $v$ or $v'$ in both the quantile and $r$ using two-dimensional interpolation. After this and other interpolations are set up, sampling the DFs is fast, achieving rates of $\approx 10^6$ points per second.

\subsection{Special cases and new ergodic and Osipkov-Merritt NFW distribution functions}

\begin{figure}
	\centering
	\includegraphics[width=\columnwidth]{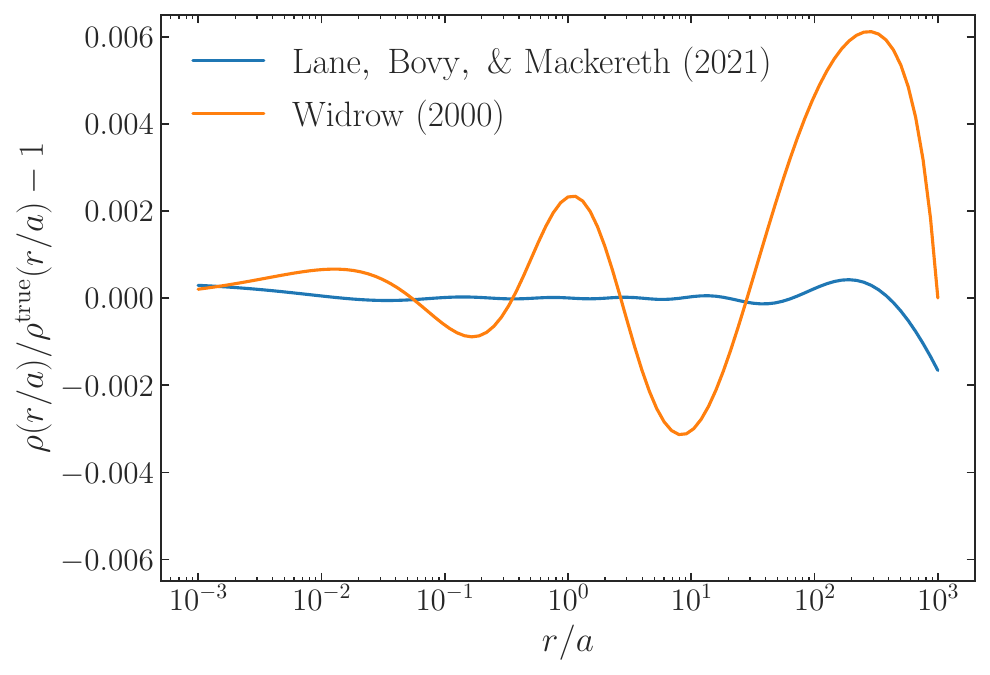}\\
	\includegraphics[width=\columnwidth]{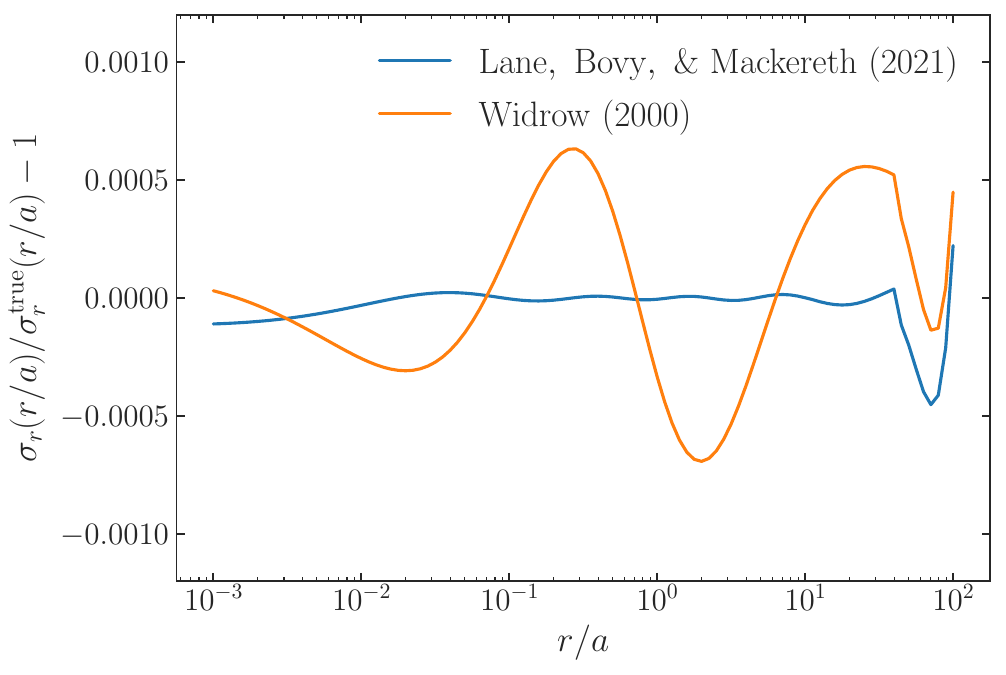}
	\caption{Recovery of the density and radial velocity dispersion of an NFW profile from the new form of Equation \eqref{eq-nfw-ergodic-newform} for its self-consistent distribution function. Compared to the simpler approximation from \citet{Widrow00a}, the match to the density and velocity dispersion is better by more than an order of magnitude almost everywhere using our improved fitting function.}
	\label{fig:ergodic_nfw}
\end{figure}

Many analytic solutions for ergodic and anisotropic spherical DFs exist. Of these we have currently implemented:
\begin{itemize}
    \item The ergodic DF of a Hernquist profile \citep{hernquist90};
    \item The Osipkov-Merritt-type DF of a Hernquist profile \citep{hernquist90};
    \item The constant $\beta$ DF for a Hernquist profile for arbitrary $\beta$ \citep{baes02};
    \item The ergodic DF of a Plummer profile $f(\mathcal{E}) \propto \mathcal{E}^{-7/2}$ \citep[e.g.,][]{binney08};
    \item The \citet{king66} DF (see also \citealt{michie63}).
\end{itemize}

Because NFW profiles describe the spatial distribution of simulated dark-matter halos well, self-consistent NFW DFs are of special interest. However, there is no known analytic form for the ergodic or general anisotropic DFs (however, see \citealt{Evans06a} for a notable exception, the DF with constant $\beta = 1/2$). Approximate ergodic and Osipkov-Merritt-type DFs for the NFW profile were presented by \citet{Widrow00a}. Using the high-fidelity numerical calculation obtained using our code of the ergodic and Osipkov-Merritt-type NFW DFs, we have found more complex, but more accurate fitting functions for these two cases. For the ergodic DF, we find that
\begin{equation}
\begin{split}\label{eq-nfw-ergodic-newform}
    f(\mathcal{E})\times&(4\pi G)^{3/2}\,a^3\,\rho_0^{1/2} \\&\!\!\!= \mathcal{E}^{3/2} \,(1-\mathcal{E})^{-5/2}\left({-\ln \mathcal{E} \over 1-\mathcal{E}}\right)^{-2.75}\\& \times \left(7.84806318891231\, \mathcal{E}^{10} - 41.0268009529576\, \mathcal{E}^{9}\right.\\& \quad\ + 92.5144063082258\, \mathcal{E}^{8} - 117.647787290798\, \mathcal{E}^{7}\\&\quad\ + 92.6397009471828\, \mathcal{E}^{6} - 46.6587221550258\, \mathcal{E}^{5}\\&\quad\ + 14.9776586391246\, \mathcal{E}^{4} - 2.97848277491979\, \mathcal{E}^{3}\\&\quad\ + 0.258346829924101\, \mathcal{E}^{2} + 0.0232272797489981\, \mathcal{E}\\&\left. \quad\ + 0.0926081086527954\right),
\end{split}
\end{equation}
where $\mathcal{E}$ is evaluated in its dimensionless form $\mathcal{E}/(4\pi\,G\,\rho_0\,a^2)$.
Figure \ref{fig:ergodic_nfw} compares the resulting density and radial-velocity dispersion profiles to their exact values, illustrating that this corresponds to the NFW profile better than the \citet{Widrow00a} approximation.

For the Osipkov-Merritt-type DF for the NFW profile, we find that 
\begin{equation}
\begin{split}
    \huge[f(Q)&-f^{\beta=0}(Q)\huge]\times(4\pi G)^{3/2}\,a^3\,\rho_0^{1/2} \\&\!\!\!=  \left({a \over r_a}\right)^2\, {1 \over Q^{2/3}}\,\left({1-Q \over \ln Q}\right)^2\,\\
    & \times \left(- 0.995895790138335 \,Q^{8} + 4.29052661245253 \,Q^{7}\right.\\&\quad\
    - 7.60690467091859\, Q^{6} + 7.03132348658788\, Q^{5}\\&\quad\ - 3.6920719890718\, Q^{4} + 0.831302363461598\, Q^{3}\\&\quad\ - 0.217968733177408\, Q^{2} - 0.0408426627412238\, Q\\&\quad\ \left.+ 0.0802975743915827\right),
\end{split}
\end{equation}
where $Q$ is evaluated in its dimensionless form $Q/(4\pi\,G\,\rho_0\,a^2)$ and $f^{\beta=0}(Q)$ is the ergodic DF from Equation \eqref{eq-nfw-ergodic-newform} evaluated at $\mathcal{E} = Q$. Unlike the forms given by \citet{Widrow00a}, our new form allows for an arbitrary value of $a/r_a$. How well the approximate DF recovers the density and radial velocity dispersion profile depends on $a/r_a$, but the density is typically recovered to within a few percent out to $100a$ and the velocity dispersion to better than one percent.


\bsp	
\label{lastpage}
\end{document}